\documentclass[aps,pra,twocolumn,showpacs,floatfix,twoside,superscriptaddress,eqsecnum]{revtex4-2}
\usepackage{amssymb}
\usepackage{graphicx}
\usepackage{amsmath}
\usepackage{amsfonts}
\usepackage{esint}
\usepackage{color}
\usepackage{xcolor}
\usepackage{float}
\usepackage{subcaption}
\usepackage{verbatim}
\usepackage{bm}
\usepackage[colorlinks=true,citecolor=blue,linkcolor=blue,urlcolor=blue]{hyperref}
\usepackage[justification=centerlast,font={small}]{caption}

\DeclareMathOperator{\sign}{sign}

\newcommand{\BT}{\mathcal{T}}

\newcommand{\BP}{\mathcal{P}}

\newcommand{\BC}{\mathcal{C}}

\begin{document}
 
 \title{Formation of Exceptional Points in pseudo-Hermitian Systems}
\author{Grigory A. Starkov}
\email{Grigorii.Starkov@rub.de}
\author{Mikhail V. Fistul}
\author{Ilya M. Eremin}
\affiliation{Institut fur Theoretische Physik III, Ruhr-Universitat Bochum, 44801 Bochum, Germany}

\date{\today}

\begin{abstract}
Motivated by the recent growing interest in the field of $\BP\BT$-symmetric 
Hamiltonian systems we theoretically study the emergency of singularities called Exceptional Points (\textit{EP}s) in the eigenspectrum of pseudo-Hermitian Hamiltonian as the strength of Hermiticity-breaking terms turns on.  Using general symmetry arguments, we characterize the separate energy levels by a topological $\mathbb{Z}_2$ index which corresponds to the signs $\pm 1$ of the eigenvalues of pseudo-metric operator $\hat \zeta$ in the absence of Hermiticity-breaking terms. After that, we show explicitly that the formation of second-order \textit{EP}s is governed by this $\mathbb{Z}_2$-index: only the pairs of levels with \textit{opposite} index can provide second-order \textit{EP}s.
Our general analysis
is accompanied by a detailed study of \textit{EP}s appearance in 
an exemplary  $\BP\BT$-symmetric pseudo-Hermitian system with parity operator in the role of $\hat \zeta$: a transverse-field Ising spin chain with a staggered imaginary longitudinal field.
Using analytically computed parity indices of all the levels, we analyze the eigenspectrum of the model in general, and the formation of third-order \textit{EP}s in particular.
\end{abstract}

\maketitle

\section{Introduction}
Quantum mechanics on a macroscopic scale fascinates scientists for many years, and recently it becomes a firm basis of various applications in the field of Quantum Information, i.e., quantum computing, quantum simulators etc. \cite{acin2018quantum}. However, the quantum mechanics applies to the isolated systems only, and for macroscopic systems interacting with an environment unavoidably present  dissipation and decoherence lead to the  degradation of coherent quantum properties, e.g., the decay of quantum beats or microwave induced Rabi oscillations with the time \cite{acin2018quantum,mooij1999josephson}. 

A one natural way to describe the dynamics of dissipative quantum systems is to use \textit{non-Hermitian }Hamiltonians. Indeed, a non-Hermitian Hamiltonian has complex eigenvalues, and e.g., an exponential decay of quantum properties that is typical for the dissipative quantum dynamics, can be easily obtained with this approach. 

Beyond that
the non-Hermitian systems have their own unique  properties. Thus, one of the main drivers of the active research of non-Hermitian systems is the emergency of the \textit{singularity points} in the eigenspectrum of a non-Hermitian Hamiltonian. Such points are called "Exceptional Points" (\textit{EP}s), and they correspond to the points in the parameter space of a non-Hermitian Hamiltonian, where two or more eigenvalues and corresponding eigenvectors coalesce, rendering the Hamiltonian non-diagonalizable~\cite{Xiao-19,Ding2022rev}.
Second-order \textit{EP}s have peculiar topological properties~\cite{Bergholz-21, HuSunChen-22}, and can be used in various applications such as quantum sensors~\cite{Wiersig-14, Wiersig-16, WiersigYang-17, Khajavikhan-19}, adiabatic time asymmetric quantum-state-exchange~\cite{Richter-04, Moiseyev-13}, mode switching in waveguides~\cite{Chong-16, Rotter-16, Chan-18, Ghosh-18} and optical microcavities~\cite{Ghosh-17, Ghosh-17b, Ghosh-19},  laser emission management~\cite{Rotter-14, XZhang-16} just to name a few.

Moreover,
higher-order \textit{EP}s, which involve three or more coalescing states, have been also theoretically studied, and there has already been a number of interesting proposals that utilize the higher-order \textit{EP}s, e.g., an increase of the sensitivity of the quantum sensors ~\cite{Khajavikhan-17},  states conversion~\cite{Ghosh-20, Ghosh-21} and  speeding up entanglement generation~\cite{Whaley-22}.

A main challenge in realization of the quantum devices based on \textit{EP}s is
the need to tune a device to the vicinity of a given \textit{EP}. The crucial step towards solution of this problem 
is to obtain general conditions determining  the formation of \textit{EP}s in non-Hermitian systems.
Classification of \textit{EP}s and quantification of the number of independent parameters required for fine-tuning of a system to the \textit{EP} have been done in Refs.~\cite{Hatsugai-21, Bergholz-21b, Sayyad-22, Kunst-22}.
The analysis carried out in these works was limited to the vicinities of \textit{EP}s and focused on the states that coalesce at corresponding \textit{EP}s.
As such, this analysis provides only a partial picture as it does not allow to identify the 
\textit{specific} states in the eigenspectrum of the Hamiltonian \textit{forming }\textit{EP}s as the Hermiticity-breaking terms are turned on.

The number of the parameters required for fine-tuning is reduced if the Hamiltonian is subject to a symmetry from the class of anti-unitary symmetries~\cite{Hatsugai-21, Bergholz-21b, Sayyad-22, Kunst-22}.
This class includes parity--time-reversal ($\BP\BT$), parity--particle-hole ($\BC\BP$), pseudo-chiral and pseudo-Hermitian (\textit{psH}) symmetries. These symmetries are not unrelated, as they all can be connected to the notion of pseudo-Hermiticity: The substitution $\hat{H}\rightarrow -i\hat{H}$ turns $\BC\BP$-symmetry into $\BP\BT$-symmetry, while pseudo-chiral symmetry --- into pseudo-Hermitian one~\cite{Hatsugai-21}.
Finally, $\BP\BT$-symmetric Hamiltonians are known to be pseudo-Hermitian~\cite{mostafazadeh2002pseudo,mostafazadeh2,mostafazadeh3,mostafazadeh2010, Ashida-20,ZhangQinXiao-20}.

A pseudo-Hermitian Hamiltonian
satisfies  the following condition:
\begin{equation} \label{PHH}
 \hat\zeta \hat {H}  = \hat{H}^\dagger \hat{\zeta},
\end{equation}
where $\hat \zeta=\hat \zeta^\dagger$ is an invertible Hermitian operator, called a
pseudo-metric operator.
This condition leads to a special symmetry of the eigenspectrum of a pseudo-Hermitian Hamiltonian:
the spectrum consists of purely real eigenvalues and pairs of complex conjugated eigenvalues,
and  the \textit{EP}s separate the regions of parameters where the levels have real and complex eigenvalues.

In this paper, we are going to consider pseudo-Hermitian Hamiltonians, for which a globally invariant pseudo-metric $\hat\zeta$ can be chosen, that does not depend on the parameters of the Hamiltonian. This case corresponds to the pseudo-Hermitian (\textit{psH}) symmetry of the Hamiltonian~\cite{Sayyad-22}.
In comparison, for a $\BP\BT$-symmetric Hamiltonian, $\hat\zeta$ is not uniquely defined~\cite{bian2020conserved,agarwal2022conserved}, and it is not apriori known whether such globally invariant choice of $\hat\zeta$ can be made.
Nevertheless, there are
plenty of examples of $\BP\BT$-symmetric systems which are simultaneously \textit{psH}-symmetric.

In the manuscript, we present a theoretical study of the formation of \textit{EP}s  in the eigenspectrum of an arbitrary \textit{psH}-symmetric Hamiltonian under the assumption that there are no additional symmetries that would lead to symmetry protected degeneracies. We relate the formation of \textit{EP}s to the topological $\mathbb{Z}_2$-index corresponding to the signs of the eigenvalues $\pm 1$ of the pseudo-metric operator $\hat \zeta$ in the absence of Hermiticity-breaking terms. 

Our general analysis is then applied to an exemplary model of
transverse-field Ising spin chain in an imaginary staggered longitudinal magnetic field, which
has been introduced by us in Ref.~\cite{StarkovFistulEremin-22} in the context of the ground state quantum phase transitions.
This model is \textit{psH}-symmetric, and the parity operator can be chosen as the globally invariant pseudo-metric: $\hat\zeta=\hat\BP$.
Interestingly enough, this model is simultaneously $\BP\BT$-symmetric.
Since the parity is self-inverse operator with $\pm1$ eigenvalues, the topological $\mathbb{Z}_2$-index coincides in this case with the parity of a state at vanishing imaginary field.
In the presence of the non-Hermitian terms, the model becomes non-integrable, however, if one turns these terms off, the model is exactly diagonalizable via the combination of the Jordan-Wigner and the generalized Bogolyubov transformations~\cite{Lieb-61}.
Using the analytically computed parities of all the states, we analyze the formation of  \textit{EP}s of second and third order in different regimes.

Such a model can be experimentally realized as an array of interacting $\BP\BT$-symmetric qubits (two-level systems), where imaginary longitudinal magnetic field corresponds to a proper combination of gain (loss).
The quantum dynamics of  $\BP\BT$-symmetric single qubit have been observed in numerous atomic or solid state systems such as trapped ions, ultracold and Rydberg atoms \cite{ding2021experimental,lourencco2022non,li2019observation}, Bose-Einstein condensate \cite{cartarius2012model}, superconducting \cite{naghiloo2019,dogra2021quantum} or  nitrogen-vacancies \cite{wu2019observation} qubits interacting with auxiliary qubits. In all these systems a non-equilibrium growth of the population of specially chosen quantum states, {\it i.e.}, states with a \textit{gain}, can be completely compensated by a \textit{loss } present in the other states, and therefore, the $PT$-symmetric combination of gain and loss occurs. Moreover, recently the entanglement generation has been studied \cite{Whaley-22}, and a general analysis have been used to identify different quantum regimes for two $PT$-symmetric  interacting qubits \cite{tetling2022linear}.

The paper is organized as follows: In Section~\ref{section:index}, we define the topological $\mathbb{Z}_2$-index and show that the eigenstates of a \textit{psH}-symmetric Hamiltonian
can be characterized by such $\mathbb{Z}_2$-index
in the regions of parameters where the corresponding eigenvalues stay real. In Section~\ref{section:rule}, we derive the selection rules determining the formation of \textit{EP}s in the eigenvalues spectrum of an arbitrary $psH$-symmetric Hamiltonian.
In Section~\ref{section:ising}, we apply our general analysis elaborated in Sec.~\ref{section:index} and~\ref{section:rule} to an exemplary \textit{psH}-symmetric non-Hermitian system: transverse field Ising chain with staggered longitudinal gain and loss, and show how the formation of second-order and third-order \textit{EP}s relates to the $\mathbb{Z}_2$-indices of eigenstates, which are equivalent to the parities in this case. 
In Section~\ref{section:conclusions}, we present the discussion of the results and conclusions.

There are two Appendices:
in Appendix~\ref{section:diagonalization}, following Ref.~\cite{Lieb-61}, we provide a recount of the properties of transverse-field Ising model; in Appendix~\ref{section:parity}, the parities of all states are computed for vanishing imaginary field.

\section{Definition of the $\mathbb{Z}_2$-index of a state.\label{section:index}}

Let us consider a \textit{psH}-symmetric Hamiltonian
$\hat{H}(\vec p)$ which depends on the vector of physical parameters $\vec p$:
\begin{equation}
    \hat\zeta \hat H(\vec p) = \hat H^\dagger(\vec p)\hat\zeta\qquad\text{for any }\vec p.\label{pph2}
\end{equation}
Varying the parameters $\vec{p}$ one can move the eigenvalues and eigenvectors of $\hat{H}(\vec p)$ closer or further to the Exception points (\textit{EP}s).

Away from \textit{EP}s, one can introduce the complete bi-orthonormal set of left and right eigenvectors of the Hamiltonian~\cite{Mostafazadeh-10}:
\begin{align}
 \hat H(\vec p) |R_n(\vec p)\rangle & = \varepsilon_n(\vec p) |R_n(\vec p)\rangle,\\
 \langle L_n(\vec p) | \hat H(\vec p) & = \varepsilon_n(\vec p)\langle L_n(\vec p)|,\label{left-right-eigvecs}
\end{align}
\begin{equation}
 \langle L_n(\vec p) | R_{n^\prime}(\vec p) \rangle = \delta_{n,n^\prime}.\label{completeness}
\end{equation}

Let us consider the right $|R_\alpha(\vec p)\rangle$ and the left $\langle L_\alpha(\vec p)|$ eigenvectors of some state $\alpha$. For real values of $\epsilon_\alpha$
we obtain by Hermitian conjugation of Eq.~\eqref{left-right-eigvecs}
\begin{equation}
    \hat{H}^\dagger(\vec p)|L_\alpha(\vec p)\rangle = \varepsilon_\alpha(\vec p)|L_\alpha(\vec p)\rangle.
\end{equation}
On the other hand, it follows from Eq.~\eqref{pph2} that
\begin{equation}
    \hat{H}^\dagger(\vec p)\hat{\zeta}|R_\alpha(\vec p)\rangle = \hat{\zeta}\hat{H}(\vec p)|R_\alpha(\vec p)\rangle = \varepsilon_\alpha(\vec p) \hat{\zeta}|R_\alpha (\vec p)\rangle.
\end{equation}
Assuming that the state $\alpha$ is not degenerate, we obtain
\begin{equation}
   |L_\alpha(\vec p)\rangle =c(\vec p)\hat{\zeta}|R_\alpha(\vec p)\rangle, \label{lpr-relation}
\end{equation}
where $c(\vec p)$ is a \textit{real-valued }
coefficient. Indeed, it
follows from the fact that
\begin{equation}
\langle R_\alpha(\vec p)|\hat{\zeta}|R_\alpha(\vec p)\rangle = c(\vec p)\langle R_\alpha(\vec p)|L_\alpha(\vec p)\rangle = c(\vec p),
\end{equation}
is the quantum average of Hermitian operator.
Notice here that the bi-orthonormal set of the left and the right eigenvectors is defined up to the arbitrary rescaling factor
providing that the equation~\eqref{completeness} is satisfied. 
It means that one can multiply $|R_\alpha(\vec p)\rangle$ by some arbitrary complex $w(\vec p)\neq 0$ function
and simultaneously divide $|L_\alpha(\vec p)\rangle$ by this $w(\vec p)$.
After this rescaling, the relation between $\hat\zeta |R_\alpha(\vec p)\rangle$ and $|L_\alpha(\vec p)\rangle$ becomes
\begin{equation}
    |L_\alpha(\vec p)\rangle = \frac{c(\vec p)}{\left|w(\vec p)\right|^2}\hat\zeta |R_\alpha(\vec p)\rangle.
\end{equation}
Choosing  $w(\vec p) = \sqrt{|\langle R_\alpha(\vec p)|\hat\zeta|R_\alpha(\vec p)\rangle|}$ we obtain
\begin{equation}
    |L_\alpha(\vec p)\rangle = \zeta_\alpha \hat\zeta |R_\alpha(\vec p)\rangle,\label{mapping}
\end{equation}
where
\begin{equation}
    \zeta_\alpha = \sign{\left[\langle R_\alpha(\vec p)|\hat{\zeta}|R_\alpha(\vec p)\rangle\right]} = \pm1\label{parity-index}
\end{equation}
is the $\mathbb{Z}_2$-index of the state $\alpha$.

As we see from Eq.~\eqref{parity-index}, the $\mathbb{Z}_2$-index $\zeta_\alpha$ can only change at the points where the $\langle R_\alpha(\vec p)|\hat{\zeta}|R_\alpha(\vec p) \rangle$ turns to zero. It takes place at an \textit{EP}, where necessarily
\begin{equation}
\langle R_\alpha(\vec p)|\hat{\zeta}|R_\alpha(\vec p) \rangle = 0.\label{EP-condition}
\end{equation}
Indeed, at an \textit{EP}, $|R_\alpha(\vec p)\rangle = |R_\beta(\vec p)\rangle$, where $\beta$ is some other state. In the vicinity of the exception point, $\hat \zeta |R_\alpha(\vec p)\rangle$ is orthogonal to $|R_\beta(\vec p)\rangle$, and it stays orthogonal at the exception point by continuity.
In comparison, at a regular point, $\hat\zeta|R_\alpha(\vec p)\rangle$ is a non-zero vector (because $\hat\zeta$ is invertible) proportional to $\hat\zeta|R_\alpha(\vec p)\rangle$, and $\langle R_\alpha(\vec p)|\hat{\zeta}|R_\alpha(\vec p) \rangle\neq 0$ because of Eq.~\eqref{completeness}.

To complete this analysis, we notice that in complex
\textit{psH}-symmetric quantum systems (see an exemplary model below, Sec.~\ref{section:ising}), there can be
the points of accidental degeneracy where the energy level $\alpha$ crosses some other level $\beta$ without forming an \textit{EP}. However, using the condition of the orthogonality of $\hat{\zeta}|R_\alpha(\vec p)\rangle$ and $|R_\beta(\vec p)\rangle$ vectors outside of the accidental degeneracy point and the continuity arguments, one can conclude that
Eq.~\eqref{lpr-relation} should also hold at the accidental degeneracy points.

To conclude this section, we have explicitly shown that the $\mathbb{Z}_2$-index of the pseudo-metric Hermitian operator $\hat \zeta$ is 
an invariant characteristic of a state of a \textit{psH}-symmetric Hamiltonian in the whole range of parameters $\vec p$ where the corresponding eigenvalue takes real values.
As a result, we can choose arbitrary parameters in this region to compute the $\mathbb{Z}_2$-index of a state. 
It is especially convenient to choose parameters in such a way, that the non-Hermitian part of the Hamiltonian is zero:
In this case the Hamiltonian commutes with $\hat\zeta$ (see Eq.~\eqref{pph2}), and the $\mathbb{Z}_2$-indices coincide with the signs of the eigenvalues of $\hat\zeta$, which characterize the common eigenstates of the Hamiltonian and pseudo-metric operator.

\section{Selection rules governing the formation of the second-order \textit{EP}s.\label{section:rule}}

In this section, we show explicitly that the two eigenstates of a \textit{psH}-symmetric Hamiltonian form a second-order \textit{EP} only if they have opposite $\mathbb{Z}_2$-indices. 
Qualitative explanation of that is following:
firstly, in the vicinity of an \textit{EP}, the two states forming \textit{EP} become very close in the energy so that the influence of other states is negligible; secondly,
at the point very close to \textit{EP} one can project the Hamiltonian onto two states forming the \textit{EP}, and then treat  small changes of parameters as the perturbation. If the states have equal $\mathbb{Z}_2$-indices, the projected Hamiltonian is locally Hermitian and does not lead to any \textit{EP}s.

To prove that, we track two eigenstates in the region of parameters $\vec p$ where the eigenvalues are real-valued.
Let $\varepsilon_i$, $|R_i(\vec p)\rangle$ and $|L_i(\vec p)\rangle$ be the eigenvalues, the right and the left eigenvectors for these states $i=1,2$, accordingly. We will also assume, that the eigenvectors has already been rescaled to satisfy Eq.~\eqref{mapping}.

Choosing the vector of parameters $\vec p^{\,\prime}$ very close to some particular \textit{EP} we
linearize 
the Hamiltonian $\hat H$ in the vicinity of 
$\vec p^{\,\prime}$ as
\begin{equation}
 \hat H(\vec p) = \hat H_0^\prime + \left(\vec p-\vec p^{\,\prime}\right)\cdot \hat{\vec{V}}\label{ham-lin}
\end{equation}
and treat the second term in r.h.s. of~\eqref{ham-lin} as the perturbation.
Here, $\hat H_0^\prime = \hat H(\vec p^{\,\prime})$, and $\hat {\vec V}$ is
\begin{equation}
 \hat {\vec V} = \left.\frac{\partial \hat H}{\partial \vec p}\right|_{\vec p=\vec p^{\,\prime}}.
\end{equation}
Since $\vec p^{\,\prime}$ is not an \textit{EP}, the Hamiltonian $\hat H_0^\prime$ has a full biorthonormal set of left and right eigenvectors. 

The eigenvalues and eigenstates of the pseudo-Hermitian Hamiltonian (\ref{ham-lin}) can be obtained similarly to the case of Hermitian Hamiltonian \cite{Dirac-1930} but 
the matrix elements $\langle m | \hat{V}|n\rangle$ have to be substituted by the matrix elements $\langle L_m | \hat{V}|R_n\rangle$.
Since the difference $\varepsilon_1-\varepsilon_2$ between the levels energies can be made arbitrary small by choosing $\vec p^{\,\prime}$ arbitrary close to the \textit{EP}, we conclude  
that only the states with energies $\varepsilon_1$ and $\varepsilon_2$ are important in the vicinity of the corresponding \textit{EP}.

Projecting the \textit{psH}-symmetric Hamiltonian~\eqref{ham-lin} on these two states we obtain the resulting projected Hamiltonian as a $2\times2$ matrix $\hat M_{i,j}(\vec p)=\langle L_i^\prime | \hat{H}(\vec p) |R_j^\prime\rangle$,
where $\{i,j\}=1,2$, $|R_j\prime\rangle = |R_j(\vec p^{\,\prime})\rangle$ and $\langle L_i^\prime| = \langle L_i(\vec p^{\,\prime})|$.
A more explicit expression for $\hat M_{i,j}(\vec p)$ is
\begin{equation}
 \hat M_{i,j}(\vec p) =
 \begin{pmatrix}
  \varepsilon_1 (\vec p^{\,\prime}) & 0\\
  0 & \varepsilon_2 (\vec p^{\,\prime})
 \end{pmatrix}
 + (\vec p-\vec p^{\,\prime})\cdot \langle L_i^\prime| \hat{\vec V} |R_j^\prime\rangle.\label{ham-lin-proj}
\end{equation}

By making use of the identity ~\eqref{mapping} and the
defining property of \textit{psH}-symmetric Hamiltonian~\eqref{pph2},
we 
relate the matrix elements of the projected Hamiltonian to the matrix elements of its Hermitian conjugate, e.g.,
\begin{multline}
 \langle L_i^\prime |\hat{\vec V}| R_j^\prime\rangle = \zeta_i \langle R_i^\prime| \hat \zeta \hat{\vec V}| R_j^\prime\rangle 
  = 
  \zeta_i \langle R_i^\prime | \hat {\vec V}^\dagger \hat \zeta | R_j^\prime\rangle = 
  \\=
  \zeta_i\zeta_j \langle R_i^\prime |\hat {\vec V}^\dagger| L_j^\prime\rangle = \zeta_i\zeta_j \overline{\langle L_j^\prime |\hat {\vec V}| R_i^\prime\rangle}.\label{conjugation}
\end{multline}
Substituting it into Eq.~\eqref{ham-lin-proj}, we obtain the explicit form of the projected Hamiltonian as
\begin{equation}
 M_{i,j}(\vec p) = \begin{pmatrix}
          \varepsilon_1(\vec p^{\,\prime}) + (\vec p-\vec p^{\,\prime})\cdot \vec u_1 & (\vec p-\vec p^{\,\prime})\cdot \vec w\\
          \zeta_1\zeta_2 (\vec p-\vec p^{\,\prime})\cdot \vec w^*   & \varepsilon_2(\vec p^{\,\prime}) + (\vec p-\vec p^{\,\prime})\cdot \vec u_2,
         \end{pmatrix},\label{projected-form}
\end{equation}
where $\vec u_1$ and $\vec u_2$ are real-valued vector-functions of $\vec p^{\,\prime}$, while $\vec w$ and $\vec w^*$ are complex conjugated vector-functions of $\vec p^{\,\prime}$.

If $\zeta_1\zeta_2=1$, the projected Hamiltonian is \textit{locally Hermitian} one, and there is no \textit{EP} in the vicinity of $\vec p^{\,\prime}$. 
Therefore, the eigenstates with the same $\mathbb{Z}_2$-index  cannot form the second-order \textit{EP}s.

\section{Case Study: Transverse-field Ising spin chain with staggered longitudinal gain and loss.\label{section:ising}}

In this section, we apply the general symmetry analysis elaborated in Sections~\ref{section:index} and~\ref{section:rule} to an exemplary system which is both \textit{psH}-symmetric and $\BP\BT$-symmetric: transverse-field Ising spin chain with staggered longitudinal gain and loss. 

\subsection{Model}
Let us consider a one-dimensional chain composed of an even number $N$ of spins-$1\over2$ in the presence of transverse magnetic field of strength $\Delta$ along $x$ direction. The adjacent spins are coupled by the Ising type interaction $-J\hat\sigma_n^z\hat\sigma_{n+1}^z$ and are subject to the staggered gain and loss, which can be modeled by the staggered imaginary longitudinal magnetic field $(-1)^{n-1}\,\mathrm{i}\gamma \hat{\sigma}_{n}^{z}$:
\begin{equation}
    \hat{H} = \sum_{n=1}^{N} \left[\Delta\hat{\sigma}_{n}^{x}+(-1)^{n-1}\,\mathrm{i}\gamma \hat{\sigma}_{n}^{z}\right]-J\sum_{n=1}^{N-1} \hat{\sigma}_{n}^{z}\hat{\sigma}_{n+1}^{z},
    \label{Ising-Ham}
\end{equation}
where a single parameter $\gamma$ is a strength of gain (loss). 

This model has been previously elaborated in the context of the ground state quantum phase transitions~\cite{StarkovFistulEremin-22}.
A similar model with complex transverse magnetic field has been studied also previously~\cite{Song-14,Song-15,Schmidt-21}.
The important difference with the case of complex transverse magnetic field is that adding the longitudinal magnetic field breaks the integrability of the transverse-field Ising chain. In Ref.~\cite{StarkovFistulEremin-22}, it has been also shown that such a system demonstrates numerious \textit{EP}s as the strength $\gamma$ increases. 

Next, we define the parity $\hat \BP$ and the time reversal $\hat\BT$ operators as the mirror reflection and complex conjugation respectively~\cite{Song-14,tetling2022linear,StarkovFistulEremin-22}:
\begin{equation}
    \hat\BP \hat\sigma_{n}^\alpha \hat\BP = \hat\sigma_{N+1-n}^\alpha,\qquad \hat\BT i \hat\BT = -i.\label{symmetry-actions}
\end{equation}
Here, we used the fact that thus defined $\hat\BP$ and $\hat\BT$ are self-inverse operators.
The Hamiltonian~\eqref{Ising-Ham} is a $\BP\BT$-symmetric one.
Moreover, it is \textit{psH}-symmetric, and
the parity operator $\hat\BP$ having eigenvalues $\pm 1$ can be used as the globally invariant pseudo-metric operator $\hat \zeta$
, i.e., $\hat \zeta=\hat \BP$. Correspondingly, the $\mathbb{Z}_2$-index coincides with parity at vanishing imaginary field becomes simply the \textit{parity-index}.

By making use of the direct numerical diagonalization of the Hamiltonian~\eqref{Ising-Ham}, we calculate the eigenspectrum of~\eqref{Ising-Ham} for a wide range of physical parameters, $\Delta$, $J$ and $\gamma$. The details of our approach can be found in \cite{StarkovFistulEremin-22}. Since for $\gamma=0$ the Hamiltonian~\eqref{Ising-Ham} becomes an integrable one, the eigenspectrum in this case can be obtained analytically (see Ref.~\cite{Lieb-61} and Appendix~\ref{section:diagonalization}). We are focusing specifically on the case of open boundary conditions, because there are no additional symmetries that would induce symmetry-protected degeneracies then.

\subsection{The eigenspectrum: second-order \textit{EP}s.}

\begin{figure*}[t]
\begin{center}
\includegraphics[width=450pt]{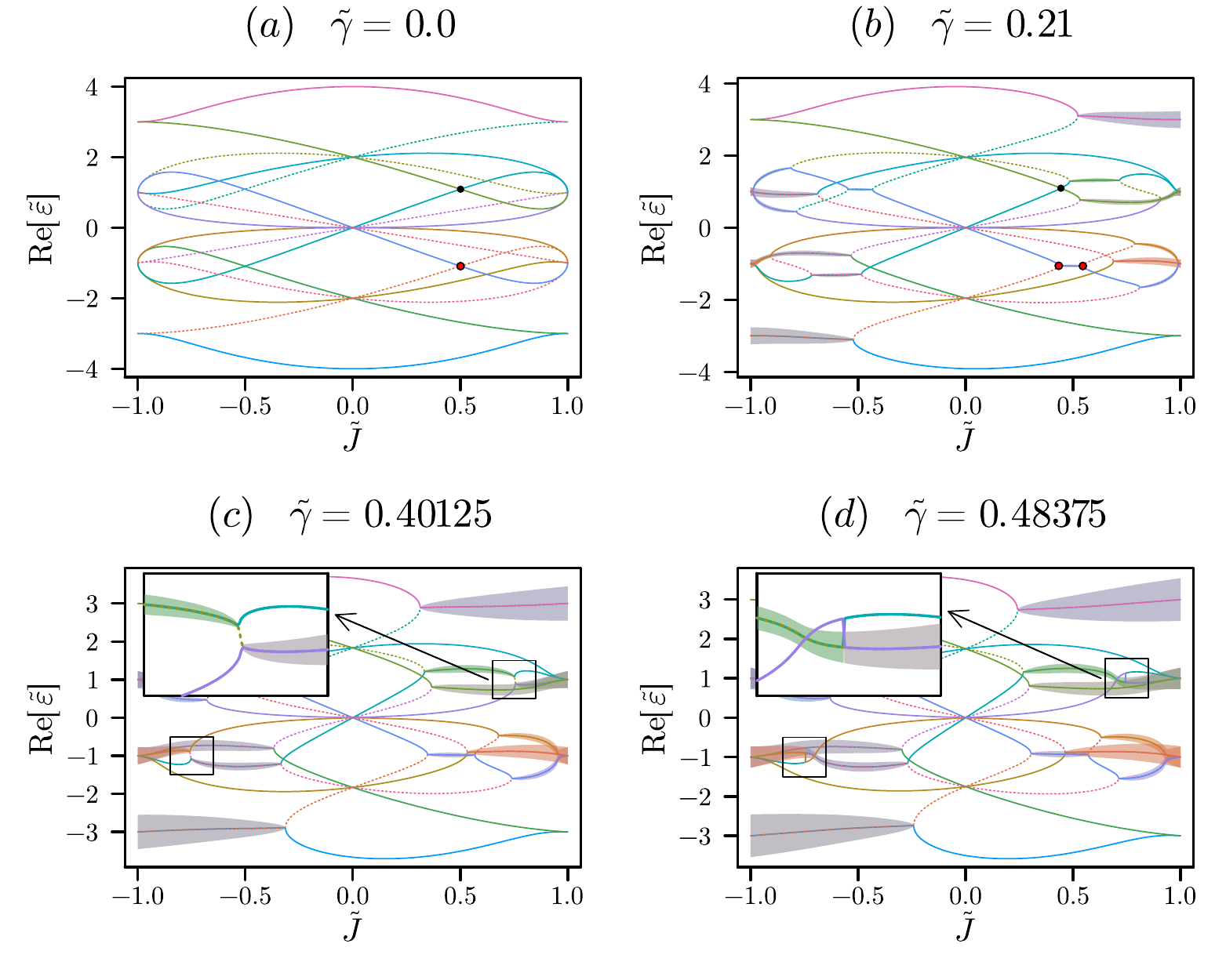}
\caption{The dependence of a real part of normalized eigenvalues $\tilde{\varepsilon}_n=\varepsilon_n/\sqrt{J^2+\Delta^2}$ on the normalized coupling strength $\tilde J=J/\sqrt{J^2+\Delta^2}$ for a $N=4$ spin chain. The parameter $\tilde\gamma=\gamma/\sqrt{J^2+\Delta^2}$ was chosen as $(a)\ 0;\ (b)\  0.21;\ (c)\ 0.40125;\ (d)\ 0.48375$. The shaded ribbons around the curves depict the scaled imaginary part of the normalized eigenvalues.
The parities of the states are denoted by differing lines styles: solid for $\BP=1$ and dotted for $\BP=-1$.
In the panels (a) and (b): the red dots show how a particular \textit{opposite parity }crossing splits into two second-order \textit{EP}s; black dots mark a stable same parity crossing.
In the panels (c) and (d), small open rectangles mark the positions of the third order exception points. The insets in the panels (c) and (d) enlarge the content of the open rectangles and display only the levels involved in the formation of a third-order \textit{EP\/}.}
\label{stp4}
\end{center}
\end{figure*}

\begin{figure*}[t]
\begin{center}
\includegraphics[width=450pt]{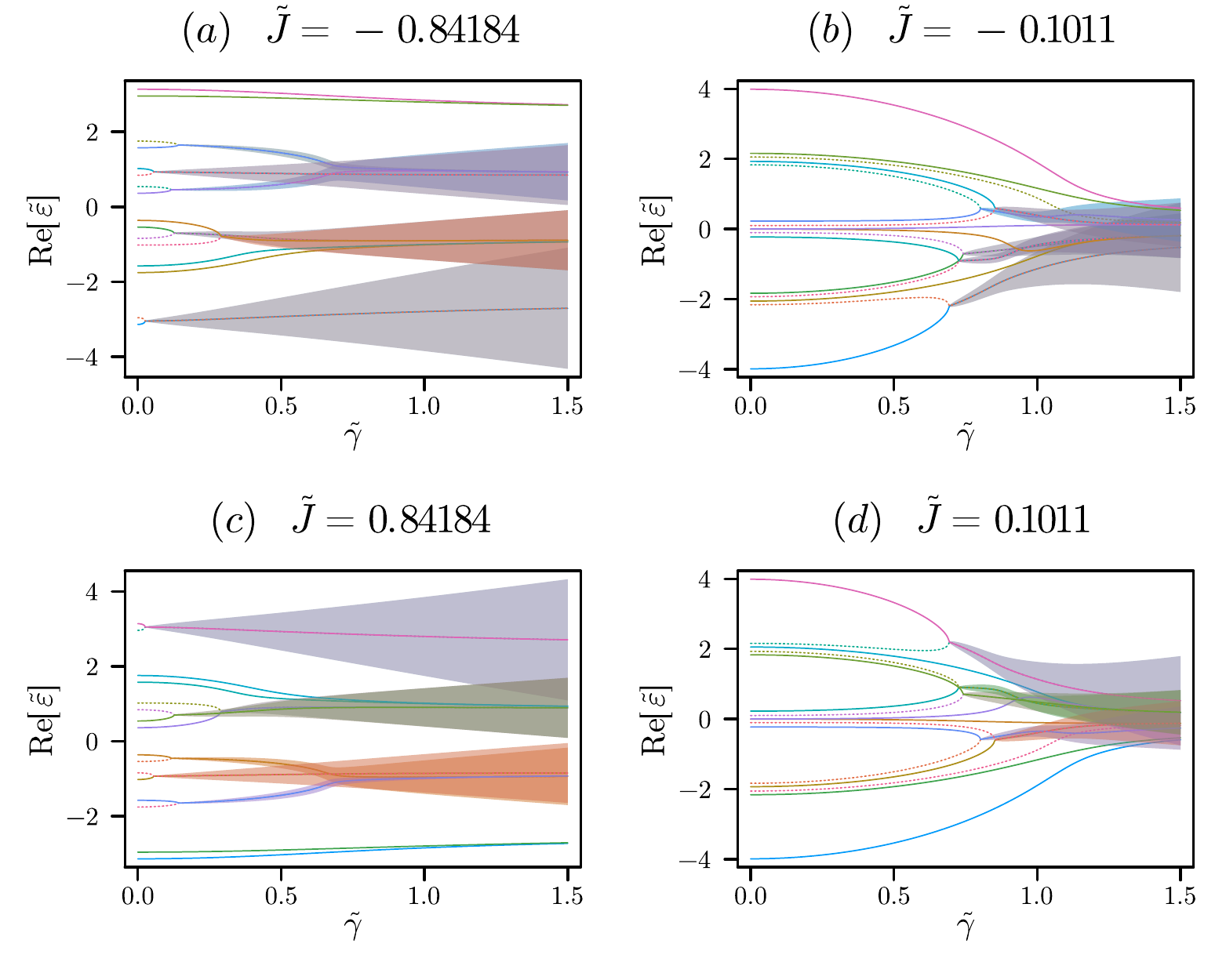}
\caption{The dependence of a real part of normalized eigenvalues $\tilde{\varepsilon}_n=\varepsilon_n/\sqrt{J^2+\Delta^2}$ on the normalized strengh of non-Hermiticity $\tilde\gamma=\gamma/\sqrt{J^2+\Delta^2}$ for a $N=4$ spin chain. The normalized coupling strength $\tilde J=J/\sqrt{J^2+\Delta^2}$ was chosen as $(a)\ -0.84184;\ (b)\ -0.1011;\ (c)\ 0.84184;\ (d)\  0.1011$. The shaded ribbons around the curves depict the scaled imaginary part of the normalized eigenvalues.
The parities of the states are denoted by differing lines styles: solid for $\BP=1$ and dotted for $\BP=-1$.
}
\label{stj4}
\end{center}
\end{figure*}

In Fig.~\ref{stp4}, we present the results of the numerical diagonalization of the Hamiltonian~\eqref{Ising-Ham} for an exemplary case of a chain of $N=4$ spins.
The four subfigures display the dependence of the normalized eigenvalues~$\tilde\varepsilon = \varepsilon/\sqrt{J^2+\Delta^2}$ on the normalized Ising coupling strength $\tilde J = J/\sqrt{J^2+\Delta^2}$ for different fixed values of $\tilde\gamma=\gamma/\sqrt{J^2+\Delta^2}$. 
A complementary picture is presented in Fig.~\ref{stj4}, where the four subfigures display the dependence of $\tilde\varepsilon$ on $\tilde \gamma$ for different fixed values of $\tilde J$.
We have computed the parities of all the states as explained in Appendix~\ref{section:parity} and denoted them visually by solid (dotted) styles of lines.
The pairs of the complex conjugated pairs are denoted by shaded ribbons. When two levels go through a second order \textit{EP}, they form a complex conjugated pair. As such, the \textit{EP}s can be identified in Figures by looking at the points where the ribbons end.
Overall, the Figs.~\ref{stp4} and~\ref{stj4} very clearly demonstrate the \textit{$\mathbb{Z}_2$-index selection rule} in action.

\subsubsection{The ground state\label{gstate-parity}}

Many of the features of the eigenspectrum are universal, {\it i.e.,\/} they do not really depend neither on the chain length $N$ 
nor on the specific distribution of the imaginary longitudinal field (we can consider some other distributions that differ from the staggered one), as long as the Hamiltonian keeps $\hat\BP$-pseudo-Hermitian symmetry.
One of such features is the formation of the \textit{EP}s by the ground and the first excited states.
For the antiferromagnetic sign of the interaction $J<0$, the ground state always goes through the \textit{E}P with the first excited state as $\tilde\gamma$ becomes sufficiently large (see Fig. \ref{stp4}b-d). At the same time, for the ferromagnetic sign of the interaction $J>0$, the ground state does not form \textit{EP} point (see Fig. \ref{stp4}b-d).
This fact has a simple explanation in terms of the parities of the states: for $J>0$, both the ground and the first excited states has positive parities, while for $J<0$, the first excited state has the negative parity.
This can be also understood from the following argument. Let us focus on the strong-coupling limit $|J|\gg \Delta$. For the ferromagnetic coupling, the ground and the first excited states are symmetric and antisymmetric combinations of spin states $|\uparrow\uparrow\dotsc\uparrow\rangle $ and
$|\downarrow\downarrow\dotsc\downarrow\rangle$
\begin{align}
    |0\rangle & = \frac{|\uparrow\uparrow\dotsc\uparrow\rangle + |\downarrow\downarrow\dotsc\downarrow\rangle}{\sqrt{2}},\label{gsf} \\
    |1\rangle & = \frac{|\uparrow\uparrow\dotsc\uparrow\rangle - |\downarrow\downarrow\dotsc\downarrow\rangle}{\sqrt{2}}.\label{fesf}
\end{align}
The parity operator (see Eq.~\eqref{symmetry-actions}) exchanges the leftmost spin with the rightmost, the next-to-leftmost spin with the next-to-rightmost, {\it etc\/.}
When we apply the parity operator $\hat \BP$ to Eqs.~\eqref{gsf} and~\eqref{fesf}, the both states $|0\rangle$ and $|1\rangle$ do not change sign.
For the antiferromagnetic coupling, however, the ground and the first excited states have the form
\begin{align}
    |0\rangle & = \frac{|\uparrow\downarrow\dotsc\uparrow\downarrow\rangle + |\downarrow\uparrow\dotsc\downarrow\uparrow\rangle}{\sqrt{2}},\label{gsaf} \\
    |1\rangle & = \frac{|\uparrow\downarrow\dotsc\uparrow\downarrow\rangle - |\downarrow\uparrow\dotsc\downarrow\uparrow\rangle}{\sqrt{2}}.\label{fesaf}
\end{align}
Although the ground state (\ref{gsaf}) still has the positive parity, the application of the parity operator to the state (\ref{fesaf}) flips the sign of the first excited state.

In Ref.~\cite{StarkovFistulEremin-22}, we have considered the ground state phase transitions in the limit $N\rightarrow \infty$ for the periodic boundary conditions. We have argued there, that the phase transition into the ordered phase always happens via a second-order \textit{EP} for the antiferromagnetic sign of the interaction, while for the ferromagnetic sign of the interaction the phase transition occurs without any \textit{EP}s and the ground state energy stays real. The argument about the parities that we have just provided applies also to the case of the periodic boundary conditions.
As a consequence, it serves as a clarification of the results of Ref.~\cite{StarkovFistulEremin-22}.

The atiferromagnetic symmetric and antisymmetric combinations~\eqref{gsaf} and~\eqref{fesaf} are the two highest-in-energy states for the ferromagnetic sign of the interaction, and {\it vice versa\/.} As such, the situation is reversed in comparison with the ground state: the two highest-in-energy states do not produce the \textit{EP}s
for the antiferromagnetic interaction, and go through a second-order \textit{EP} for the ferromagnetic interaction (see Fig. \ref{stp4}b-d).

\subsubsection{Strong-coupling limit $|J|\gg\Delta$.\label{section:strong-coupling}}
At $\gamma=0$, the Hamiltonian~\eqref{Ising-Ham} 
is integrable, and it can be diagonalized analytically via the combination of Jordan-Wigner and generalized Bogolyubov transformations (see Appendix~\ref{section:diagonalization} and Ref.~\cite{Lieb-61}). After diagonalization, the Hamiltonian~\eqref{Ising-Ham}
takes the form:
 \begin{equation}
  \hat H_0 = -\frac12\sum_{i=0}^{N-1} \varepsilon_{k_i} + \sum_{i=0}^{N-1} \varepsilon_{k_i}\eta_{k_i}^\dagger\eta_{k_i},
 \end{equation}
where $\eta_{k_i}^\dagger$ and $\eta_{k_i}$  are the creation and annihilation operators of Bogolyubov-transformed fermionic modes.

In the strong-coupling limit, there are $N-1$ modes that form a band of width $\sim \Delta$ with the energies $\approx |J|$. There is also one almost zero fermionic mode with the exponentially small energy
\begin{equation}
 \varepsilon_{k_0} \approx 2|J|\left(1 - \frac{\Delta^2}{J^2}\right)\left(\frac{\Delta}{|J|}\right)^N.\label{almost-zero}
\end{equation}
Overall, the eigenspectrum is split into well isolated bands with the fixed number of excited non-zero fermionic modes.
At the same time all the states form almost degenerate pairs. The two states in a pair differ by the filling of almost zero $k_0$ fermionic mode and the energy splitting is given by Eq.~\eqref{almost-zero}.

Using Eqs.~\eqref{recursive-parity-raw} and~\eqref{delta-general}, we can find the relative parity of the two levels forming an almost degenerate pair:
\begin{equation}
 \zeta_{-k_0}\zeta_{+k_0} = \sign\left[J\right](-1)^r,
\end{equation}
where $r$ is the number of excited non-zero fermionic modes.
The labels $-k_0$ and $+k_0$ refer to ``unoccupied" and ``occupied" $k_0$-mode respectively. 
For the case of the ferromagnetic coupling, almost degenerate pairs have opposite parities in every odd-$r$ band and identical parities in every even-$r$ band.
For the case of the antiferromagnetic coupling, the situation is opposite: almost degenerate pairs have opposite parities in every even-$r$ band and identical parities in every odd-$r$ band (For the ground state one can put $r=0$.)

Let us focus on one of the bands where the almost degenerate pairs have opposite parity.
The almost degenerate pairs are well separated: the distance between different almost degenerate pairs is of the order $\Delta$, while the splitting inside the pair is exponentially small (see  Eq.~\eqref{almost-zero}).
When considering a single almost degenerate pair, the influence of all the other states is negligible, and, similarly to the
Section~\ref{section:rule}, we expand  the Hamiltonian at $\gamma=0$ and obtain the matrix $\hat M_{i,j}(\gamma)$ as $\gamma$ is turned on:
\begin{equation}
    \hat M = \varepsilon_{J,\Delta}
    \hat{\mathrm{I}}
    +
    \begin{pmatrix}
        \varepsilon_{k_0}/2 & \gamma w\\
        -\gamma w^*           & -\varepsilon_{k_0}/2
    \end{pmatrix}
    \label{adp-form}
\end{equation}
where $\hat{\mathrm{I}}$ is $2\times2$ identity matrix and
\begin{multline}
    w = \langle +k_0| \frac{\partial \hat{H}}{\partial \gamma}|-k_0\rangle
    = \\
    \langle +k_0| i\sum_{n=1}^{N-1} (-1)^{n-1}\hat\sigma_n^z|-k_0\rangle
\end{multline}

Comparing the general Eq.~\eqref{projected-form} with Eq.~\eqref{adp-form}, we note that $\vec p-\vec p^{\,\prime}$ corresponds to $\gamma$, while $\varepsilon_{1,2}(\vec p^{\,\prime})=\varepsilon_{J,\Delta}\pm\varepsilon_{k_0}/2$. The diagonal matrix elements $u_1$ and $u_2$ from Eq.~\eqref{projected-form} are equal to zero in Eq.~\eqref{adp-form} because
the diagonal matrix elements of the anti-Hermitian operator $\hat V = \partial \hat H/\partial \gamma$ at $\gamma=0$ are equal to zero.

The two eigenvalues of $\hat{M}$ are
\begin{equation}
\varepsilon_\pm = \varepsilon_{J,\Delta}\pm\sqrt{\varepsilon_{k_0}^2/4 - \gamma^2|w|^2}.
\end{equation}
As we see, if the off-diagonal matrix element $w$ is not accidentally zero, the gap closes and the almost degenerate pair goes through a second-order \textit{EP} at the critical value of $\gamma_{cr}$ which is linear in gap $\varepsilon_{k_0}$:
\begin{equation}
    \gamma_{cr} = \frac{\varepsilon_{k_0}}{2|w|}.
\end{equation}
For larger values of $\gamma$, the eigenvalues form a complex conjugated pair. Since the gap
$\varepsilon_{k_0}$ is strongly diminished in the limit of large $J$ and/or $N$, the second-order \textit{EP}s appear already at extremely small values of $\gamma$.

Let us compare it with Figs.~\ref{stj4}$(a,c)$. Here,
the strong-coupling limit corresponds to $|\tilde J|\approx 1$. The levels $3$--$8$ form the first excited band while the levels $9$-$14$ form the second excited band. In the first excited band of the ferromagnet (panel $(c)$ ) and the second excited band of the antiferromagnet (panel $(a)$) the almost degenerate pairs quickly go through second-order \textit{EP}s and form complex conjugated pairs in accordance to our analysis.

In the second excited band of the ferromagnet and the first excited band of the antiferromagnet, the second-order \textit{EP}s are formed only between the levels from different almost degenerate pairs, and some of the states do not form \textit{EP}s at all
(see states $13$,$14$ in ferromagnet, $3$,$4$ --- in antiferromagnet).

\subsubsection{Weak-coupling limit $|J|\ll\Delta$\label{section:weak-coupling}}

In this case, all the Bogolyubov-transformed fermionic modes form a band of width $\sim|J|$ with the energies $\approx \Delta$. So, all the eigenspectrum is divided into the bands with different total number of excited fermionic modes. However, there are no direct matrix elements $w$ of the anti-Hermitian part of the Hamiltonian between the states of the same band: The terms contributing to the anti-Hermitian part of the Hamiltonian consist of the odd number of fermionic creation and annihilation operators (see Eq.~\eqref{gain-loss-jw}). But the states  with the same number of excited Bogolyubov-transformed fermionic modes are connected by an even number of fermionic creation and annihilation operators.

As a result, the second-order \textit{EP}s are formed by the pairs of states belonging to different bands (see Figs.~\ref{stj4}$(b,d)$).
Another consequence is that the eigenspectrum is overall more robust towards the formation of \textit{EP}s.

\subsubsection{Intermediate coupling: stability of level crossings.\label{section:crossings}}

As we go from the weak-coupling limit to the strong-coupling limit at $\gamma=0$, the levels reorder themselves and the energy level crossings occur (see Fig.~\ref{stp4}(a)). Let us discuss what happens with these crossings as $\gamma$ is turned on.

At the crossing, the pair of levels is well isolated in energy from all the other levels, so yet again we apply the approach of Sections~\ref{section:rule} and~\ref{section:strong-coupling}.
Let us focus on the case of the opposite parities of the crossing levels (red dot in Fig.~\ref{stp4}(a) is an example).
At $\gamma=0$, we can write the projected Hamiltonian in the vicinity of the crossing point as
\begin{equation}
\hat M = \varepsilon_{\Delta J} \hat{\mathrm{I}} + \alpha\Delta J  \hat\sigma_z,
\end{equation}
where $\hat{\mathrm I}$ is $2\times2$ identity matrix
and parameter $\Delta J$ measures the detuning from the crossing point.
When we add small $\gamma$, the projected Hamiltonian becomes (see Eqs.~\eqref{projected-form} and~\eqref{adp-form})
\begin{equation}
  \hat M = \varepsilon_{\Delta J} \hat{\mathrm{I}} +
  \begin{pmatrix}
    \alpha \Delta J & \gamma w\\
    -\gamma w^* & -\alpha \Delta J
  \end{pmatrix}
\end{equation}
The two eigenvalues in this case are
\begin{equation}
    \varepsilon_\pm = \varepsilon_{\Delta J} \pm \sqrt{\alpha^2(\Delta J)^2 - \gamma^2|w|^2};
\end{equation}
At finite $\gamma$, the crossing is split into two second-order \textit{EP}s at $|\Delta J| = \gamma |w|/\alpha$.
For $|\Delta J| > \gamma |w|/\alpha$ the eigenvalues are real, while in between the two \textit{EP}s $|\Delta J| < \gamma |w|/\alpha$, the two eigenvalues form a complex conjugated pair.

This is precisely what happens with the opposite parity crossings in Fig.~\ref{stp4}.
As an example, we have singled out one of such crossings in panel $(a)$ and marked it with a red dot.
In panel $(b)$, we have marked the two corresponding \textit{EP}s after splitting by red dots as well.

In the case of same parity crossing, the transformation of the exact crossing point into the \textit{EP} is strictly forbidden. 
Moreover, the off-diagonal matrix element $w$ at $\gamma=0$ is equal to zero, i.e., in linear in $\gamma$ order, the gap at the same parity crossing does not open.
However, when we go to finite $\gamma$, the general form of the projected Hamiltonian~\eqref{projected-form} does not exclude the possibility of generation of off-diagonal matrix elements. If such matrix elements were to appear, it would lead to avoided crossings and opening of the gap~\cite{NeumannWigner-29,grifoni1998driven}.

It is noteworthy, that in the model we considered, the same parity crossings are stable and do not open gaps.
As an example, we have marked one of such crossings with a black dot in Figs.~\ref{stp4}$(a,b)$.
We argue that it is a consequence of the special structure of the anti-Hermitian part of the Hamiltonian~\eqref{Ising-Ham}: as we have mentioned in Section~\ref{section:weak-coupling}, it consists only of the terms that are odd in the number of fermionic operators. We have experimented with the different distributions of the imaginary field.
For example, we have considered a staggered imaginary transverse field (Song's model~\cite{Song-14}) and a staggered imaginary field with both transverse and longitudinal components.
In the former case, the anti-Hermitian part of the Hamiltonian consists only of even in fermionic operators terms, and the same parity states form stable exact crossings.
In the latter case, however, we saw the gaps opening because of avoided crossing. Overall, this topic voids further exploration in a future work.

\subsection{The eigenspectrum: third-order \textit{EP}s.\label{section:ep3}}

\begin{figure*}[t]
\includegraphics[width=390pt]{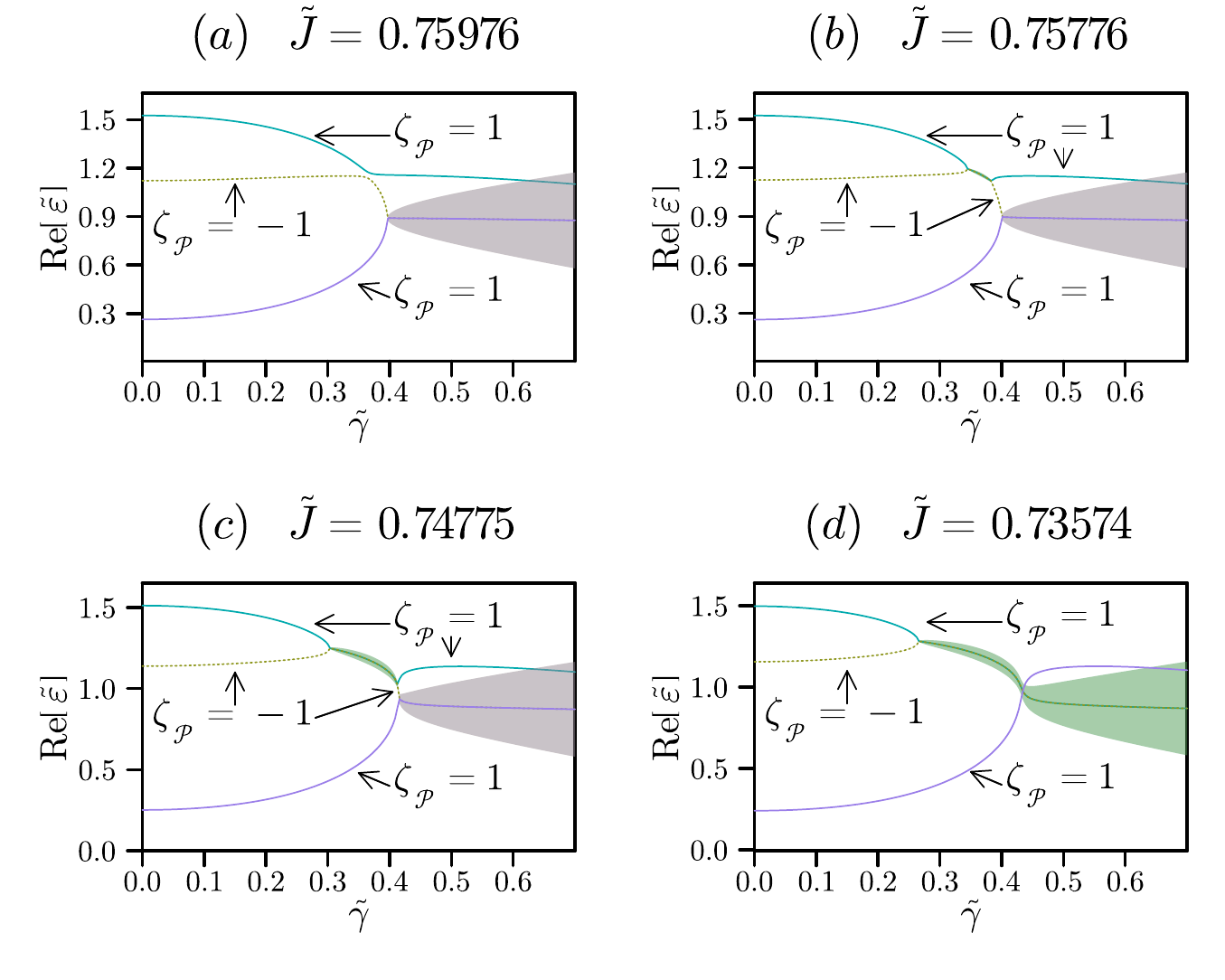}
\caption{The structure of the exception point of the third order. The plots show the real part of the normalized eigenvalues as functions of the normalized imaginary field strength $\tilde\gamma = \gamma/\sqrt{J^2+\Delta^2}$ for different fixed values of normalized coupling constant $\tilde J = J/\sqrt{J^2 + \Delta^2}$.
The shaded ribbons depict the scaled imaginary part of the normalized eigenvalues.
Only the levels forming the exception point are plotted. The annotations specify the parities of the states.}
\label{ep3}
\end{figure*}

For larger values of $\tilde\gamma$, the two exception points of the second order can coalesce and go through the exception point of the third order (\textit{EP}3)~\cite{Somnath-20,Bergholtz-21} as seen in Fig.~\ref{stp4}(c,d) and insets therein. To clarify the structure of \textit{EP}3, we provide the complementary picture in Fig.~\ref{ep3}, where we display the normalized eigenvalues of the levels forming the highlighted \textit{EP}3 as the functions of $\tilde\gamma$ for different values of $\tilde J$. The parities of the levels are denoted both by annotations and by differing line styles. The \textit{EP}3 is formed by three nearby levels, where the lower and the upper levels have the same parity index, while the middle level has the opposite parity index. The middle level is allowed to form a complex conjugated pair with either of the other two levels, while it is disallowed by selection rule for the lower and the upper level. As a result, the vicinity of \textit{EP}3 is characterized by the competition between these two possible pairings of the middle level. For example, for larger values of $\tilde{J}$ (see Fig.~\ref{ep3}(a)), the upper level's eigenvalue stays real, while the middle and the lower levels form a complex conjugated pair. On the contrary, for smaller values of $\tilde{J}$ (see Fig.~\ref{ep3}(d)) the situation is opposite. An interesting consequence of \textit{EP}3, is that for larger values of $\tilde{\gamma}$, there is an abrupt exchange between the upper and lower levels as $\tilde{J}$ is varied (compare panels (c) and (d) of Fig.~\ref{ep3}). This is consistent with the jumps visible in the insets of Fig.~\ref{stp4}(c,d) and is attributed to the branching nature of the singularity at \textit{EP}3.

When the value $\tilde\gamma$ is increased even further, the system passes through an increasing number of third-order \textit{EP}s at different values of $\tilde J$. In this section, we focused on a particular third-order \textit{EP\/}. All the other third-order \textit{EP}s that we have observed in the system, have the same structure that we have just described: coalescing pair of second-order \textit{EP}s with shared level of opposite parity.

\section{Discussion and Conclusion\label{section:conclusions}}

In this paper, we have theoretically studied the formation of Exceptional Points (\textit{EP}s) in the eigenpectrum of a Hamiltonian with pseudo-Hermitian (\textit{psH}) symmetry
as the strength of Hermiticity-breaking terms increases.

We have shown, that for an overall non-degenerate level with real eigenvalue, the left and the right eigenvectors can always be rescaled to satisfy
\begin{equation}
    |L_n\rangle = \zeta_n \hat{\zeta}|R_n\rangle,
\end{equation}
where
\begin{equation}
    \zeta_n = \sign\left[\langle R_n |\hat{\zeta}|R_n\rangle\right] = \pm1.
\end{equation}
The $\mathbb{Z}_2$-index $\zeta_n$ is conserved in the whole region of parameters where the level has real eigenvalue and coincides with the signs of the eigenvalues of the pseudometric operator $\hat\zeta$ when anti-Hermitian part of the Hamiltonian is zero.
After that, we have shown that  the formation of a second-order \textit{EP} is governed by index-based selection rule: the formation of \textit{EP} is impossible for a pair of levels with identical $\mathbb{Z}_2$-indices.

To clearly illustrate these ideas, we have considered a transverse-field Ising spin chain with imaginary staggered longitudinal field, which has pseudo-Hermitian symmetry with respect to the parity operator.
In this case, the $\mathbb{Z}_2$ index coincides with the parity of a state at vanishing imaginary field.
The system is integrable in the absence of Hermiticity-breaking terms, which enabled us to compute the parities of all the levels analytically.
Using the knowledge of the level parities, we have applied the selection rule to analyse second-order \textit{EP} formation in the strong- and the weak-coupling limits, as well as the stablity of level crossings with respect to Hermiticity-breaking terms.
It is noteworthy, that in the considered model, the selection rule for \textit{EP} formation is very similar to the selection rule that governs avoided crossing, although the condition on the parities is reversed~\cite{NeumannWigner-29}: the avoided crossing happens when the parities are identical, while \textit{EP} formation --- when they are opposite.

We have also demonstrated, how the third-order \textit{EP}s can be obtained by tuning two second-order \textit{EP}s with shared level of opposite $\mathbb{Z}_2$-index to coincide.
In principle, a fourth-order \textit{EP} can be obtained by tuning two third-order \textit{EP}s with shared levels, a fifth-order \textit{EP} --- by tuning two fourth-order \textit{EP} with shared levels, {\it etc.}.
As a consequence of the selection rule, a higher-order \textit{EP} in a \textit{psH}-symmetric system must be formed by the levels with {\it staggered} signature of $\mathbb{Z}_2$-indices.

Moreover, the analysis of the paper goes beyond the specific shape of the anti-Hermitian term we considered.
In principle, it stays relevant even for general form of anti-Hermitian terms, as long as they preserve pseudo-Hermitian symmetry. This is attributed to the fact, that the parity signature of the levels significantly limits the possible structure of the eigenspectrum.

For example, one could consider a more general distribution of the imaginary longitudinal field:
\begin{equation}
 \hat{H}_{ah} = \mathrm{i}\sum_{n=1}^{N} \gamma_n \hat\sigma_n^z
\end{equation}
This introduces additional independent control parameters which is necessary to tune the system to higher-order \textit{EP}s.
If we require that $\gamma_n = -\gamma_{N+1-n}$, then the Hamiltonian stays both $\BP\BT$-symmetric and $\hat\BP$-\textit{psH}-symmetric.

In Section~\ref{section:index}, we did not really specify the nature of the parameters $\vec p$. In principle, one can consider $\hat H (\vec p)$ to be a Bloch Hamiltonian of some extended system, where the vector of parameters $\vec {p}$ contains the Bloch wave-vector as part of its components. It follows from the fact that pseudo-Hermitian symmetry does not change the wave-vector~\cite{Sayyad-22}, so it acts as the internal symmetry of the Bloch Hamiltonian. It is potentially interesting to apply the analysis of the paper to the non-Hermitian topological systems.

Let us finally mention that the instabilities of \textit{psH}-symmetric Hamiltonians were systematically studied in 50-s by the Soviet mathematicians Krein~\cite{Krein-1950} and Gel'fand with Lidskii~\cite{GelfandLidskii-1955} in the context of linear differential equations with periodic coefficients.
A consistent recollection of their works can be found in the book~\cite{StarzhinskiiYakubovich-1975}.
They defined a pseudo-scalar product
\begin{equation}
 \langle\chi,\psi\rangle = \langle\chi|\hat \zeta|\psi\rangle
\end{equation}
which was then used to classify all the real eigenvalues of the Hamiltonian as first, second and mixed kind depending on the signature of the restriction of this pseudo-scalar product to the corresponding \textit{eigensubspace}: positive definite, negative definite and undefinite respectively. After that they have shown, that the eigenvalues of the first and second kind are stable with respect to the perturbations of the Hamiltonian, while the mixed kind eigenvalues can branch off into complex conjugated pairs.

In the case of generally non-degenerate levels, the conditions on the eigenvalues of the first and second kind are equivalent to the positive or negative $\mathbb{Z}_2$-index (see~\eqref{parity-index}), while the condition on the eigenvalue of the mixed kind is equivalent to our condition~\eqref{EP-condition} on the Exceptional Points.

\section*{Acknowledgements}
We acknowledge the financial support of Deutsche Forschungsgemeinschaft (Projekt~EF~11/10-1) and the financial support through the European Union’s Horizon 2020 research and innovation program under grant agreement No 863313 'Supergalax'.


\appendix

\section{Exact Diagonalization via Jordan-Wigner transformation.\label{section:diagonalization}}

Introducing the raising and the lowering operators $\hat\sigma_n^\pm=(\hat\sigma^y_n\pm i\hat\sigma_n^z)/2$, we define the Jordan-Wigner fermionic operators:
\begin{equation}
 c_n^{\dagger} = \prod_{j=1}^{n-1} \left(-\hat\sigma_j^x\right)\times \hat\sigma_n^{+},\quad c_n = \prod_{j=1}^{n-1} \left(-\hat\sigma_j^x\right)\times \hat\sigma_n^-.\label{jw-operators}
\end{equation}
Inverting this definition, we can write the spin operators as
\begin{align}
    \hat\sigma_n^x & = 2c_n^\dagger c_n - 1,\nonumber\\
    \hat\sigma_n^z & = -i(\hat\sigma_n^+ - \hat\sigma_n^-) = -i\prod_{j=1}^{n-1} (c_j^\dagger+c_j)(c_j^\dagger-c_j)\times (c_n^\dagger - c_n).\label{jw-inverse}
\end{align}
Substituting it into the Hamiltonian~\eqref{Ising-Ham} with open boundary conditions, we find
\begin{align}
    \hat{H} & = \hat{H}_0 + \gamma \hat{V}_\mathrm{gl},\label{ham-jw-full}\\
    \hat{H}_0 & = \Delta\sum_{n=1}^N (2c_n^\dagger c_n-1) - \nonumber \\
    & J\sum_{n=1}^{N-1}\left[c_n^\dagger c_{n+1} + c_{n+1}^\dagger c_n - c_n^\dagger c_{n+1}^\dagger - c_{n+1} c_{n}\right],\label{ham-jw}\\
    \hat{V}_\mathrm{gl} & = \sum_{n=1}^N (-1)^{n-1}\prod_{j=1}^{n-1} (c_j^\dagger+c_j)(c_j^\dagger-c_j)\times (c_n^\dagger - c_n),\label{gain-loss-jw}
\end{align}

As we see, the terms contributing to the non-Hermitian part of the Hamiltonian $\gamma \hat V_\mathrm{gl}$ contain odd numbers of fermionic operators ranging from 1 operator ($n=1$ term) to $2N+1$ operators ($n=N$ term). As a result, it is impossible to exactly daigonalize the Hamiltonian~\eqref{ham-jw-full} at $\gamma\neq 0$. 

For $\gamma=0$, the Hamiltonian $\hat H = \hat{H}_0$ and the quadratic form over the fermionic operators can be diagonalized via a generalized Bogolyubov transformation~\cite{Lieb-61}:
\begin{equation}
    \hat H_0 = -\frac12\sum_k \varepsilon_k + \sum_k \varepsilon_k \eta_k^\dagger \eta_k,\label{bogolyubov-ham}
\end{equation}
where $\eta_k^\dagger$ and $\eta_k$ are the new fermionic creation and annihilation operators:
\begin{align}
    \eta_k^\dagger & = \sum_{n=1}^N \left[\frac{\phi_{kn} + \psi_{kn}}{2}c_n^\dagger + \frac{\phi_{kn}-\psi_{kn}}{2}c_n\right],\label{etak-dagger}\\
    \eta_k & = \sum_{n=1}^N \left[\frac{\phi_{kn} + \psi_{kn}}{2}c_n + \frac{\phi_{kn}-\psi_{kn}}{2}c_n^\dagger\right].\label{etak}
\end{align}

The energies $\varepsilon_k$ and the vectors $\vec\phi_k\equiv\phi_{kn}$ and $\vec\psi_k\equiv\psi_{kn}$ are determined as the eigenvalues and the eigenvectors of certain matrices, composed from the coefficients of the Hamiltonian. Here, we don't go through the details, and only list the final results for this variables. We can refer the interested reader to Appendix A and Section IIID of~\cite{Lieb-61}.

\begin{figure}[t]
\begin{center}
 \includegraphics[width=\columnwidth]{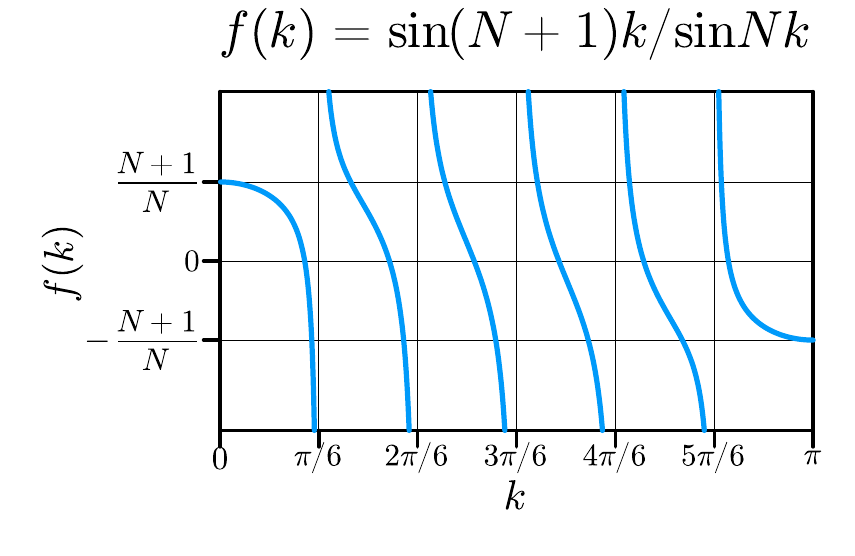}
 \caption{Plot of the function $f(k) = {\sin{(N+1)k}}/{\sin{N k}}$ for $N=6$. The equation~\eqref{eigenmode-eq} can be solved graphically by finding the intersections of the horizontal line $y = J/\Delta$ with the curve $y=f(k)$.}
 \label{fig:kmodes}
 \end{center}
\end{figure}

The energies $\varepsilon_k$ are
\begin{equation}
    \varepsilon_k = 2\sqrt{(J-\Delta)^2 + 4J\Delta\sin^2{\frac{k}{2}}},\label{eigenmode-energy}
\end{equation}
where the wave vectors $k$ are determined as the solutions of the equation
\begin{equation}
    \frac{\sin{(N+1)k}}{\sin{N k}} = \frac{J}{\Delta}.\label{eigenmode-eq}
\end{equation}
The complete set of $k$ corresponds to the interval $k\in[0,\pi]$. We display the plot of the function $f(k) = {\sin{(N+1)k}}/{\sin{N k}}$ for $N=6$ in Fig.~\ref{fig:kmodes}. The equation~\eqref{eigenmode-eq} can be solved graphically by finding the intersections of the horizontal line $y = J/\Delta$ with the curve $y=f(k)$.

The vectors $\vec\phi_k$ and $\vec\psi_k$ have the form
\begin{equation}
    \vec{\phi_k} = A_k
    \begin{pmatrix}
        \sin{k}\\
        \sin{2k}\\
        \vdots \\
        \sin{Nk}
    \end{pmatrix}, \quad
    \vec{\psi}_k = \delta_k A_k
    \begin{pmatrix}
        \sin Nk\\
        \sin (N-1)k \\
        \vdots \\
        \sin k
 \end{pmatrix}\label{phi-psi-form}
\end{equation}
Here, $A_k$ are the normalization factors which are of no importance to us. The factors $\delta_k=\pm1$ determine the parities of the Bogolyubov-transformed fermionic modes and are equal to
\begin{equation}
    \delta_k = \sign\left[\frac{\sin{k}}{\sin{Nk}}\right].\label{def:delta}
\end{equation}

There is a symmetry between the fermionic modes energies in the cases of ferromagnet and antiferromagnet. As the left-hand side of Eq.~\eqref{eigenmode-eq} satisfies $f(k) = -f(\pi - k)$,
if $k$ is the wave vector of the fermionic mode in the case of the ferromagnet, then $\pi - k$ is the wave vector of the fermionic mode in the case of antiferromagnet. From Eq.~\eqref{eigenmode-energy} if follows then that the energies of these modes are identical.

Overall, it is convenient to label the wave vectors as $k_0,\ k_1,\ \dotsc,\ k_{N-1}$ in the order of the increasing energy.
In the case of ferromagnet, it means that (consider graphical solution of Eq.~\eqref{eigenmode-eq} with the help of Fig.~\ref{fig:kmodes})
\begin{equation}
k_i\in \left[\frac{\pi}{N}i,\frac{\pi}{N}(i+1)\right],\label{franges}
\end{equation}
while in the case of antiferromagnet it implies
\begin{equation}
    k_i\in \left[\frac{\pi}{N}(N-i-1), \frac{\pi}{N}(N-i)\right].\label{afranges}
\end{equation}
Notice, that as we move from ferromagnet to antiferromagnet, the order of the modes is reversed, which is reflected in Fig.~\eqref{stp4}(a).

Finally, we should discuss the properties of the $k_0$ mode, which is responsible for the paramagnet-(anti)ferromagnet transition in the tranverse-field Ising chain. For $|J|/\Delta > (N+1)/N$, the wave-vector $k_0$ turns complex:
\begin{align}
    k_0 &= i\kappa,\quad J>0,\label{fzeromode}\\
    k_0 &= \pi - i\kappa,\quad J<0,\label{afzeromode}
\end{align}
where $\kappa$ satisfies
\begin{equation}
    \frac{\sinh{(N+1)\kappa}}{\sinh{N\kappa}} = \frac{|J|}{\Delta}.
\end{equation}
The corresponding energy decays exponentially with the system size for large $N$:
\begin{equation}
    \varepsilon_{k_0}\approx 2|J|\left(1-\frac{\Delta^2}{J^2}\right)\left(\frac{\Delta}{|J|}\right)^N.
\end{equation}
For $|J|>\Delta$, the ground state becomes doubly degenerate in the thermodynamic limit which corresponds to the two degenerate vacua in the $\mathbb Z_2$-broken phase.

\section{Determination of the parities of all the states at $\gamma=0$.\label{section:parity}}

In this appendix, we are going to show how the parities of all the states can be obtained in the closed form in the absence of imaginary longitudinal magnetic field. In order to do that, we are going to build upon the notions introduced in~\ref{section:diagonalization}. The reader can also skip the derivation straight to~\ref{sec:parity-summary} to see the final results.

\subsection{Recursion formula.}

The action of the parity symmetry on the spin operators is defined as (see Eq.~\eqref{symmetry-actions})
\begin{equation}
    \hat\BP\hat\sigma_n^\pm\hat\BP = \hat\sigma_{N+1-n}^\pm,\qquad
    \hat\BP\hat\sigma_n^x\hat\BP = \hat\sigma_{N+1-n}^x.
\end{equation}
For the Jordan-Wigner fermionic creation and annihilation operators (see Eq.~\eqref{jw-operators}), however, the action of the parity symmetry is complicated by the fact, that the string of the spin operators gets reattached to the right end of the chain:
\begin{equation}
    \hat\BP \left(\prod_{j=1}^{n-1}(-\hat\sigma_j^x)\right) \hat\sigma_n^\pm \hat\BP =
    \hat\sigma_{N+1-n}^\pm \prod_{j=N+2-n}^{N} (-\hat\sigma_j^x).\label{jw-operators-symmetry}
\end{equation}

In order to reattach the spin string back to the left end of the chain, let us introduce the full spin string operator:
\begin{equation}
    \mathfrak{P} = \prod_{n=1}^N (-\hat\sigma_n^x) = \prod_{n=1}^{N}(1-2c_n^\dagger c_n).
\end{equation}
The square of $\mathfrak{P}$ is an identity operator, so let us multiply Eq.~\eqref{jw-operators-symmetry} on the right with an identity operator:
\begin{multline}
 \sigma_{N+1-n}^\pm \prod_{j=N+2-n}^N (-\sigma_j^x) = \sigma_{N+1-n}^\pm \left(\prod_{j=N+2-n}^N (-\sigma_n^x)\right)
 \times \\
 \left(\prod_{j^\prime=1}^N (-\sigma_{j^\prime}^x)\right) \mathfrak P =
 \pm \left(\prod_{j=1}^{N-n}(-\sigma_j^x)\right) \sigma_{N+1-n}^\pm \mathfrak{P},
\end{multline}
where we have used the identity
\begin{equation}
 \sigma_{N+1-n}^\pm(-\sigma_{N+1-n}^z) = \pm \sigma_{N+1-n}^\pm.
\end{equation}
This way, we can define the action of parity symmetry on bare fermionic creation and annihilation operators as
\begin{equation}
 \hat\BP c_n^\dagger\hat\BP =  c_{N+1-n}^\dagger\mathfrak{P},\qquad \hat\BP c_n \hat\BP  =  - c_{N+1-n}\mathfrak{P}.
\end{equation}

Substituting these identities into Eq.~\eqref{etak-dagger}, we find the action of the parity symmetry on the creation operator of Bogolyubov-transformed fermionic mode:
\begin{widetext}
\begin{multline}
 \hat\BP\eta_k^\dagger\hat\BP  = \sum_{n=1}^N \left[\frac{\phi_{kn} + \psi_{kn}}{2}c_{N+1-n}^\dagger - \frac{\phi_{kn}-\psi_{kn}}{2}c_{N+1-n}\right]\mathfrak{P} = \\ =
 \sum_{n=1}^N \left[\frac{\phi_{k,N+1-n} + \psi_{k,N+1-n}}{2}c_n^\dagger - \frac{\phi_{k,N+1-n}-\psi_{k,N+1-n}}{2}c_n\right]\mathfrak{P} = \\ =
 \delta_k \sum_{n=1}^{N}\left[ \frac{\phi_{kn}+\psi_{kn}}{2}c_n^\dagger + \frac{\phi_{kn}-\psi_{kn}}{2}c_n\right]\mathfrak{P} = \delta_k \eta_k^\dagger \mathfrak{P}.
\end{multline}
\end{widetext}
In the penultimate equality, we have used the identity
\begin{equation}
 \phi_{k,N+1-n} = \delta_k \psi_{kn},\qquad \psi_{k,N+1-n} = \delta_k\phi_{kn},
\end{equation}
which follows directly from the explicit form of the vectors $\vec\phi_k$ and $\vec\psi_k$ (see Eq.~\eqref{phi-psi-form}).
We can also repeat the same steps for the annihilation operator to find
\begin{equation}
 \hat\BP\eta_k\hat\BP = -\delta_k\eta_k \mathfrak{P}.
\end{equation}

With the knowledge of the transformation properties of the creation operator $\eta_k^\dagger$, the parities of all the states can be found by induction.
Each state is completely determined by the set of occupied Bogolyubov-transformed fermionic modes. Let $|\alpha\rangle$ be a state with $r_\alpha$ excited fermionic modes and parity $\zeta_\alpha$. Let us also assume that the mode with the wave vector $k$ is unoccupied in the state $\alpha$. Then, if we add this mode to the state, we obtain
\begin{multline}
    \hat\BP\eta_k^\dagger|\alpha\rangle = \hat\BP\eta_k^\dagger\hat\BP\times\hat\BP|\alpha\rangle =
    \\
    \delta_k\eta_k^\dagger\mathfrak{P}\times \zeta_\alpha|\alpha\rangle=
    (-1)^{r_\alpha-1}\delta_k \eta_k^\dagger |\alpha\rangle.\label{recursive-parity-raw}
\end{multline}

The spin string operator $\mathfrak{P}$ is equal to $+1$ for the states with an even number of original Jordan-Wigner fermions, while it is $-1$ for the states with an odd number of original Jordan-Wigner fermions. The eigenstates of the transverse-field Ising Hamiltonian~\eqref{bogolyubov-ham} mix the states with the different number of original Jordan-Wigner fermions. However, these eigenstates has definite evenness/oddness.
For example, the ground state $|0\rangle$ is even. (This can be confirmed by directly computing $\langle\Omega|\mathfrak{P}|0\rangle$ and showing that it is equal to $1$. See~\cite{Lieb-61} for the details of such a copmutation.) As such, $\mathfrak{P}|\alpha\rangle = (-1)^{r_\alpha-1}|\alpha\rangle$, which we used in the last equality of Eq.~\eqref{recursive-parity-raw}.

\subsection{Parity factors of Bogolyubov-transformed fermionic modes.}

As we see, in order to compute the parities, we need to determine the factors $\delta_k$ first. If $k\in \left(\frac{\pi}{N}i, \frac{\pi}{N}(i+1)\right)$, then $\sign{[\sin{Nk}]} = (-1)^i$. Correspondingly, if $k\in\left(\frac{\pi}N (N-i-1), \frac{\pi}{N}(N-i)\right)$, then $\sign{[\sin{Nk}]} = (-1)^{N-1-i}$. Substituting it into Eq.~\eqref{def:delta}, we find
\begin{equation}
    \delta_{k_i} = \sign\left[\frac{\sin{k_i}}{\sin{Nk_i}}\right]=(-1)^i\label{fdelta}
\end{equation}
in the case of ferromagnet and
\begin{equation}
    \delta_{k_i} = (-1)^{N-1-i}.\label{afdelta}
\end{equation}
in the case of antiferromagnet.
Here, we have used the labeling of the Bogolyubov-transformed fermionic modes introduced in~\ref{section:diagonalization} and Eqs.~\eqref{franges} and~\eqref{afranges}.

If the wave-vector $k_0$ of the lowest energy mode turns complex, then in the case of ferromagnet, we have
\begin{equation}
    \frac{\sin{k_0}}{\sin{Nk_0}} = \frac{\sinh{\kappa}}{\sinh{N\kappa}}>0\Rightarrow \delta_{k_0} = 1,
\end{equation}
while in the case of antiferromagnet,
\begin{equation}
    \frac{\sin{k_0}}{\sin{Nk_0}} = (-1)^{N-1}\frac{\sinh{\kappa}}{\sinh{N\kappa}}\Rightarrow \delta_{k_0} = (-1)^{N-1}.
\end{equation}
Here, we have used Eqs.~\eqref{fzeromode} and~\eqref{afzeromode}. Note that the resulting factors $\delta_{k_0}$ are consistent with Eqs.~\eqref{fdelta} and~\eqref{afdelta}.
Overall, we can combine Eqs.~\eqref{fdelta} and~\eqref{afdelta} together to get
\begin{equation}
    \delta_{k_i} = (\sign{J})^{N-1}(-1)^{i}.\label{delta-general}
\end{equation}

\subsection{Final results.\label{sec:parity-summary}}

Let us consider the state $\alpha$ with $r_\alpha$ excited Bogolyubov-transformed modes, which have wave-vectors $k_{i_1},\ k_{i_2},\ \dotsc,\ k_{i_{r_\alpha}}$:
\begin{equation}
    |\alpha\rangle = \prod_{l=1}^{r_\alpha} \eta_{k_{i_l}}^\dagger |0\rangle.
\end{equation}
As we have discussed in Section~\ref{gstate-parity} of the main text, the ground state has even parity.
By repeatedly applying Eq.~\eqref{recursive-parity-raw} to the ground state, we find the parity of the state $|\alpha\rangle$:
\begin{equation}
    \zeta_\alpha = \prod_{l=1}^{r_\alpha} (-1)^{l-1}\delta_{k_{i_l}} = (-1)^{\frac{r_\alpha(r_\alpha-1)}{2}}\prod_{l=1}^{r_\alpha}\delta_{k_{i_l}},
\end{equation}
where the factors $\delta_{k_i}$ are given by Eq.~\eqref{delta-general}.

We can also use Eq.~\eqref{recursive-parity-raw} to determine the relative sign of the parities of the two states. For example, let us consider the ground state $|0\rangle$ and the first excited state $|1\rangle$. The first excited state is obtained by exciting the lowest energy Bogolyubov-transoformed fermionic mode:
\begin{equation}
    |1\rangle = \eta_{k_0}^\dagger |0\rangle.
\end{equation}
Using Eqs.~\eqref{recursive-parity-raw} and~\eqref{delta-general}, we find for a lattice with even $N$
\begin{equation}
    \zeta_0\zeta_1 = \sign{J}.
\end{equation}

Analogously, we can consider the next-to-highest $|2^N-2\rangle$ and the highest $|2^N-1\rangle$ levels:
\begin{equation}
    |2^N-1\rangle = \eta_{k_0}^\dagger |2^N-2\rangle.
\end{equation}
Since the next-to-highest level $|2^N-2\rangle$ has $N-1$ excited Bogolyubov-transformed fermionic modes,
\begin{equation}
    \zeta_{2^N-2}\zeta_{2^N-1} = (-1)^{N-1}\sign{J} = -\sign{J}.
\end{equation}


\begin{thebibliography}{59}%
\makeatletter
\providecommand \@ifxundefined [1]{%
 \@ifx{#1\undefined}
}%
\providecommand \@ifnum [1]{%
 \ifnum #1\expandafter \@firstoftwo
 \else \expandafter \@secondoftwo
 \fi
}%
\providecommand \@ifx [1]{%
 \ifx #1\expandafter \@firstoftwo
 \else \expandafter \@secondoftwo
 \fi
}%
\providecommand \natexlab [1]{#1}%
\providecommand \enquote  [1]{``#1''}%
\providecommand \bibnamefont  [1]{#1}%
\providecommand \bibfnamefont [1]{#1}%
\providecommand \citenamefont [1]{#1}%
\providecommand \href@noop [0]{\@secondoftwo}%
\providecommand \href [0]{\begingroup \@sanitize@url \@href}%
\providecommand \@href[1]{\@@startlink{#1}\@@href}%
\providecommand \@@href[1]{\endgroup#1\@@endlink}%
\providecommand \@sanitize@url [0]{\catcode `\\12\catcode `\$12\catcode
  `\&12\catcode `\#12\catcode `\^12\catcode `\_12\catcode `\%12\relax}%
\providecommand \@@startlink[1]{}%
\providecommand \@@endlink[0]{}%
\providecommand \url  [0]{\begingroup\@sanitize@url \@url }%
\providecommand \@url [1]{\endgroup\@href {#1}{\urlprefix }}%
\providecommand \urlprefix  [0]{URL }%
\providecommand \Eprint [0]{\href }%
\providecommand \doibase [0]{https://doi.org/}%
\providecommand \selectlanguage [0]{\@gobble}%
\providecommand \bibinfo  [0]{\@secondoftwo}%
\providecommand \bibfield  [0]{\@secondoftwo}%
\providecommand \translation [1]{[#1]}%
\providecommand \BibitemOpen [0]{}%
\providecommand \bibitemStop [0]{}%
\providecommand \bibitemNoStop [0]{.\EOS\space}%
\providecommand \EOS [0]{\spacefactor3000\relax}%
\providecommand \BibitemShut  [1]{\csname bibitem#1\endcsname}%
\let\auto@bib@innerbib\@empty
\bibitem [{\citenamefont {Ac{\'\i}n}\ \emph {et~al.}(2018)\citenamefont
  {Ac{\'\i}n}, \citenamefont {Bloch}, \citenamefont {Buhrman}, \citenamefont
  {Calarco}, \citenamefont {Eichler}, \citenamefont {Eisert}, \citenamefont
  {Esteve}, \citenamefont {Gisin}, \citenamefont {Glaser}, \citenamefont
  {Jelezko} \emph {et~al.}}]{acin2018quantum}%
  \BibitemOpen
  \bibfield  {author} {\bibinfo {author} {\bibfnamefont {A.}~\bibnamefont
  {Ac{\'\i}n}}, \bibinfo {author} {\bibfnamefont {I.}~\bibnamefont {Bloch}},
  \bibinfo {author} {\bibfnamefont {H.}~\bibnamefont {Buhrman}}, \bibinfo
  {author} {\bibfnamefont {T.}~\bibnamefont {Calarco}}, \bibinfo {author}
  {\bibfnamefont {C.}~\bibnamefont {Eichler}}, \bibinfo {author} {\bibfnamefont
  {J.}~\bibnamefont {Eisert}}, \bibinfo {author} {\bibfnamefont
  {D.}~\bibnamefont {Esteve}}, \bibinfo {author} {\bibfnamefont
  {N.}~\bibnamefont {Gisin}}, \bibinfo {author} {\bibfnamefont {S.~J.}\
  \bibnamefont {Glaser}}, \bibinfo {author} {\bibfnamefont {F.}~\bibnamefont
  {Jelezko}}, \emph {et~al.},\ }\bibfield  {title} {\bibinfo {title} {The
  quantum technologies roadmap: a european community view},\ }\href@noop {}
  {\bibfield  {journal} {\bibinfo  {journal} {New Journal of Physics}\ }\textbf
  {\bibinfo {volume} {20}},\ \bibinfo {pages} {080201} (\bibinfo {year}
  {2018})}\BibitemShut {NoStop}%
\bibitem [{\citenamefont {Mooij}\ \emph {et~al.}(1999)\citenamefont {Mooij},
  \citenamefont {Orlando}, \citenamefont {Levitov}, \citenamefont {Tian},
  \citenamefont {Van~der Wal},\ and\ \citenamefont
  {Lloyd}}]{mooij1999josephson}%
  \BibitemOpen
  \bibfield  {author} {\bibinfo {author} {\bibfnamefont {J.}~\bibnamefont
  {Mooij}}, \bibinfo {author} {\bibfnamefont {T.}~\bibnamefont {Orlando}},
  \bibinfo {author} {\bibfnamefont {L.}~\bibnamefont {Levitov}}, \bibinfo
  {author} {\bibfnamefont {L.}~\bibnamefont {Tian}}, \bibinfo {author}
  {\bibfnamefont {C.~H.}\ \bibnamefont {Van~der Wal}},\ and\ \bibinfo {author}
  {\bibfnamefont {S.}~\bibnamefont {Lloyd}},\ }\bibfield  {title} {\bibinfo
  {title} {Josephson persistent-current qubit},\ }\href@noop {} {\bibfield
  {journal} {\bibinfo  {journal} {Science}\ }\textbf {\bibinfo {volume}
  {285}},\ \bibinfo {pages} {1036} (\bibinfo {year} {1999})}\BibitemShut
  {NoStop}%
\bibitem [{\citenamefont {Xiao}\ \emph {et~al.}(2019)\citenamefont {Xiao},
  \citenamefont {Zhang}, \citenamefont {Hang},\ and\ \citenamefont
  {Chan}}]{Xiao-19}%
  \BibitemOpen
  \bibfield  {author} {\bibinfo {author} {\bibfnamefont {Y.-X.}\ \bibnamefont
  {Xiao}}, \bibinfo {author} {\bibfnamefont {Z.-Q.}\ \bibnamefont {Zhang}},
  \bibinfo {author} {\bibfnamefont {Z.~H.}\ \bibnamefont {Hang}},\ and\
  \bibinfo {author} {\bibfnamefont {C.~T.}\ \bibnamefont {Chan}},\ }\bibfield
  {title} {\bibinfo {title} {Anisotropic exceptional points of arbitrary
  order},\ }\href {https://doi.org/10.1103/PhysRevB.99.241403} {\bibfield
  {journal} {\bibinfo  {journal} {Phys. Rev. B}\ }\textbf {\bibinfo {volume}
  {99}},\ \bibinfo {pages} {241403} (\bibinfo {year} {2019})}\BibitemShut
  {NoStop}%
\bibitem [{\citenamefont {Ding}\ \emph {et~al.}(2022)\citenamefont {Ding},
  \citenamefont {Fang},\ and\ \citenamefont {Ma}}]{Ding2022rev}%
  \BibitemOpen
  \bibfield  {author} {\bibinfo {author} {\bibfnamefont {K.}~\bibnamefont
  {Ding}}, \bibinfo {author} {\bibfnamefont {C.}~\bibnamefont {Fang}},\ and\
  \bibinfo {author} {\bibfnamefont {G.}~\bibnamefont {Ma}},\ }\bibfield
  {title} {\bibinfo {title} {Non-hermitian topology and exceptional-point
  geometries},\ }\href {https://doi.org/10.1038/s42254-022-00516-5} {\bibfield
  {journal} {\bibinfo  {journal} {Nature Reviews Physics}\ }\textbf {\bibinfo
  {volume} {4}},\ \bibinfo {pages} {745} (\bibinfo {year} {2022})}\BibitemShut
  {NoStop}%
\bibitem [{\citenamefont {Bergholtz}\ \emph {et~al.}(2021)\citenamefont
  {Bergholtz}, \citenamefont {Budich},\ and\ \citenamefont
  {Kunst}}]{Bergholz-21}%
  \BibitemOpen
  \bibfield  {author} {\bibinfo {author} {\bibfnamefont {E.~J.}\ \bibnamefont
  {Bergholtz}}, \bibinfo {author} {\bibfnamefont {J.~C.}\ \bibnamefont
  {Budich}},\ and\ \bibinfo {author} {\bibfnamefont {F.~K.}\ \bibnamefont
  {Kunst}},\ }\bibfield  {title} {\bibinfo {title} {Exceptional topology of
  non-hermitian systems},\ }\href
  {https://doi.org/10.1103/RevModPhys.93.015005} {\bibfield  {journal}
  {\bibinfo  {journal} {Rev. Mod. Phys.}\ }\textbf {\bibinfo {volume} {93}},\
  \bibinfo {pages} {015005} (\bibinfo {year} {2021})}\BibitemShut {NoStop}%
\bibitem [{\citenamefont {Hu}\ \emph {et~al.}(2022)\citenamefont {Hu},
  \citenamefont {Sun},\ and\ \citenamefont {Chen}}]{HuSunChen-22}%
  \BibitemOpen
  \bibfield  {author} {\bibinfo {author} {\bibfnamefont {H.}~\bibnamefont
  {Hu}}, \bibinfo {author} {\bibfnamefont {S.}~\bibnamefont {Sun}},\ and\
  \bibinfo {author} {\bibfnamefont {S.}~\bibnamefont {Chen}},\ }\bibfield
  {title} {\bibinfo {title} {Knot topology of exceptional point and
  non-hermitian no-go theorem},\ }\href
  {https://doi.org/10.1103/PhysRevResearch.4.L022064} {\bibfield  {journal}
  {\bibinfo  {journal} {Phys. Rev. Res.}\ }\textbf {\bibinfo {volume} {4}},\
  \bibinfo {pages} {L022064} (\bibinfo {year} {2022})}\BibitemShut {NoStop}%
\bibitem [{\citenamefont {Wiersig}(2014)}]{Wiersig-14}%
  \BibitemOpen
  \bibfield  {author} {\bibinfo {author} {\bibfnamefont {J.}~\bibnamefont
  {Wiersig}},\ }\bibfield  {title} {\bibinfo {title} {Enhancing the sensitivity
  of frequency and energy splitting detection by using exceptional points:
  Application to microcavity sensors for single-particle detection},\ }\href
  {https://doi.org/10.1103/PhysRevLett.112.203901} {\bibfield  {journal}
  {\bibinfo  {journal} {Phys. Rev. Lett.}\ }\textbf {\bibinfo {volume} {112}},\
  \bibinfo {pages} {203901} (\bibinfo {year} {2014})}\BibitemShut {NoStop}%
\bibitem [{\citenamefont {Wiersig}(2016)}]{Wiersig-16}%
  \BibitemOpen
  \bibfield  {author} {\bibinfo {author} {\bibfnamefont {J.}~\bibnamefont
  {Wiersig}},\ }\bibfield  {title} {\bibinfo {title} {Sensors operating at
  exceptional points: General theory},\ }\href
  {https://doi.org/10.1103/PhysRevA.93.033809} {\bibfield  {journal} {\bibinfo
  {journal} {Phys. Rev. A}\ }\textbf {\bibinfo {volume} {93}},\ \bibinfo
  {pages} {033809} (\bibinfo {year} {2016})}\BibitemShut {NoStop}%
\bibitem [{\citenamefont {Chen}\ \emph {et~al.}(2017)\citenamefont {Chen},
  \citenamefont {Özdemir}, \citenamefont {Zhao}, \citenamefont {Wiersig},\
  and\ \citenamefont {Yang}}]{WiersigYang-17}%
  \BibitemOpen
  \bibfield  {author} {\bibinfo {author} {\bibfnamefont {W.}~\bibnamefont
  {Chen}}, \bibinfo {author} {\bibfnamefont {{\c{S}}.~K.}\ \bibnamefont
  {Özdemir}}, \bibinfo {author} {\bibfnamefont {G.}~\bibnamefont {Zhao}},
  \bibinfo {author} {\bibfnamefont {J.}~\bibnamefont {Wiersig}},\ and\ \bibinfo
  {author} {\bibfnamefont {L.}~\bibnamefont {Yang}},\ }\bibfield  {title}
  {\bibinfo {title} {Exceptional points enhance sensing in an optical
  microcavity},\ }\href {https://doi.org/10.1038/nature23281} {\bibfield
  {journal} {\bibinfo  {journal} {Nature}\ }\textbf {\bibinfo {volume} {548}},\
  \bibinfo {pages} {192} (\bibinfo {year} {2017})}\BibitemShut {NoStop}%
\bibitem [{\citenamefont {Hokmabadi}\ \emph {et~al.}(2019)\citenamefont
  {Hokmabadi}, \citenamefont {Schumer}, \citenamefont {Christodoulides},\ and\
  \citenamefont {Khajavikhan}}]{Khajavikhan-19}%
  \BibitemOpen
  \bibfield  {author} {\bibinfo {author} {\bibfnamefont {M.~P.}\ \bibnamefont
  {Hokmabadi}}, \bibinfo {author} {\bibfnamefont {A.}~\bibnamefont {Schumer}},
  \bibinfo {author} {\bibfnamefont {D.~N.}\ \bibnamefont {Christodoulides}},\
  and\ \bibinfo {author} {\bibfnamefont {M.}~\bibnamefont {Khajavikhan}},\
  }\bibfield  {title} {\bibinfo {title} {Non-hermitian ring~laser gyroscopes
  with enhanced sagnac sensitivity},\ }\href
  {https://doi.org/10.1038/s41586-019-1780-4} {\bibfield  {journal} {\bibinfo
  {journal} {Nature}\ }\textbf {\bibinfo {volume} {576}},\ \bibinfo {pages}
  {70} (\bibinfo {year} {2019})}\BibitemShut {NoStop}%
\bibitem [{\citenamefont {Dembowski}\ \emph {et~al.}(2004)\citenamefont
  {Dembowski}, \citenamefont {Dietz}, \citenamefont {Gr\"af}, \citenamefont
  {Harney}, \citenamefont {Heine}, \citenamefont {Heiss},\ and\ \citenamefont
  {Richter}}]{Richter-04}%
  \BibitemOpen
  \bibfield  {author} {\bibinfo {author} {\bibfnamefont {C.}~\bibnamefont
  {Dembowski}}, \bibinfo {author} {\bibfnamefont {B.}~\bibnamefont {Dietz}},
  \bibinfo {author} {\bibfnamefont {H.-D.}\ \bibnamefont {Gr\"af}}, \bibinfo
  {author} {\bibfnamefont {H.~L.}\ \bibnamefont {Harney}}, \bibinfo {author}
  {\bibfnamefont {A.}~\bibnamefont {Heine}}, \bibinfo {author} {\bibfnamefont
  {W.~D.}\ \bibnamefont {Heiss}},\ and\ \bibinfo {author} {\bibfnamefont
  {A.}~\bibnamefont {Richter}},\ }\bibfield  {title} {\bibinfo {title}
  {Encircling an exceptional point},\ }\href
  {https://doi.org/10.1103/PhysRevE.69.056216} {\bibfield  {journal} {\bibinfo
  {journal} {Phys. Rev. E}\ }\textbf {\bibinfo {volume} {69}},\ \bibinfo
  {pages} {056216} (\bibinfo {year} {2004})}\BibitemShut {NoStop}%
\bibitem [{\citenamefont {Gilary}\ \emph {et~al.}(2013)\citenamefont {Gilary},
  \citenamefont {Mailybaev},\ and\ \citenamefont {Moiseyev}}]{Moiseyev-13}%
  \BibitemOpen
  \bibfield  {author} {\bibinfo {author} {\bibfnamefont {I.}~\bibnamefont
  {Gilary}}, \bibinfo {author} {\bibfnamefont {A.~A.}\ \bibnamefont
  {Mailybaev}},\ and\ \bibinfo {author} {\bibfnamefont {N.}~\bibnamefont
  {Moiseyev}},\ }\bibfield  {title} {\bibinfo {title} {Time-asymmetric
  quantum-state-exchange mechanism},\ }\href
  {https://doi.org/10.1103/PhysRevA.88.010102} {\bibfield  {journal} {\bibinfo
  {journal} {Phys. Rev. A}\ }\textbf {\bibinfo {volume} {88}},\ \bibinfo
  {pages} {010102} (\bibinfo {year} {2013})}\BibitemShut {NoStop}%
\bibitem [{\citenamefont {Ghosh}\ and\ \citenamefont {Chong}(2016)}]{Chong-16}%
  \BibitemOpen
  \bibfield  {author} {\bibinfo {author} {\bibfnamefont {S.~N.}\ \bibnamefont
  {Ghosh}}\ and\ \bibinfo {author} {\bibfnamefont {Y.~D.}\ \bibnamefont
  {Chong}},\ }\bibfield  {title} {\bibinfo {title} {Exceptional points and
  asymmetric mode conversion in quasi-guided dual-mode optical waveguides},\
  }\bibfield  {journal} {\bibinfo  {journal} {Scientific Reports}\ }\textbf
  {\bibinfo {volume} {6}},\ \href {https://doi.org/10.1038/srep19837}
  {10.1038/srep19837} (\bibinfo {year} {2016})\BibitemShut {NoStop}%
\bibitem [{\citenamefont {Doppler}\ \emph {et~al.}(2016)\citenamefont
  {Doppler}, \citenamefont {Mailybaev}, \citenamefont {Böhm}, \citenamefont
  {Kuhl}, \citenamefont {Girschik}, \citenamefont {Libisch}, \citenamefont
  {Milburn}, \citenamefont {Rabl}, \citenamefont {Moiseyev},\ and\
  \citenamefont {Rotter}}]{Rotter-16}%
  \BibitemOpen
  \bibfield  {author} {\bibinfo {author} {\bibfnamefont {J.}~\bibnamefont
  {Doppler}}, \bibinfo {author} {\bibfnamefont {A.~A.}\ \bibnamefont
  {Mailybaev}}, \bibinfo {author} {\bibfnamefont {J.}~\bibnamefont {Böhm}},
  \bibinfo {author} {\bibfnamefont {U.}~\bibnamefont {Kuhl}}, \bibinfo {author}
  {\bibfnamefont {A.}~\bibnamefont {Girschik}}, \bibinfo {author}
  {\bibfnamefont {F.}~\bibnamefont {Libisch}}, \bibinfo {author} {\bibfnamefont
  {T.~J.}\ \bibnamefont {Milburn}}, \bibinfo {author} {\bibfnamefont
  {P.}~\bibnamefont {Rabl}}, \bibinfo {author} {\bibfnamefont {N.}~\bibnamefont
  {Moiseyev}},\ and\ \bibinfo {author} {\bibfnamefont {S.}~\bibnamefont
  {Rotter}},\ }\bibfield  {title} {\bibinfo {title} {Dynamically encircling an
  exceptional point for asymmetric mode switching},\ }\href
  {https://doi.org/10.1038/nature18605} {\bibfield  {journal} {\bibinfo
  {journal} {Nature}\ }\textbf {\bibinfo {volume} {537}},\ \bibinfo {pages}
  {76} (\bibinfo {year} {2016})}\BibitemShut {NoStop}%
\bibitem [{\citenamefont {Zhang}\ \emph {et~al.}(2018)\citenamefont {Zhang},
  \citenamefont {Wang}, \citenamefont {Hou},\ and\ \citenamefont
  {Chan}}]{Chan-18}%
  \BibitemOpen
  \bibfield  {author} {\bibinfo {author} {\bibfnamefont {X.-L.}\ \bibnamefont
  {Zhang}}, \bibinfo {author} {\bibfnamefont {S.}~\bibnamefont {Wang}},
  \bibinfo {author} {\bibfnamefont {B.}~\bibnamefont {Hou}},\ and\ \bibinfo
  {author} {\bibfnamefont {C.~T.}\ \bibnamefont {Chan}},\ }\bibfield  {title}
  {\bibinfo {title} {Dynamically encircling exceptional points: In situ control
  of encircling loops and the role of the starting point},\ }\href
  {https://doi.org/10.1103/PhysRevX.8.021066} {\bibfield  {journal} {\bibinfo
  {journal} {Phys. Rev. X}\ }\textbf {\bibinfo {volume} {8}},\ \bibinfo {pages}
  {021066} (\bibinfo {year} {2018})}\BibitemShut {NoStop}%
\bibitem [{\citenamefont {Laha}\ \emph {et~al.}(2018)\citenamefont {Laha},
  \citenamefont {Biswas},\ and\ \citenamefont {Ghosh}}]{Ghosh-18}%
  \BibitemOpen
  \bibfield  {author} {\bibinfo {author} {\bibfnamefont {A.}~\bibnamefont
  {Laha}}, \bibinfo {author} {\bibfnamefont {A.}~\bibnamefont {Biswas}},\ and\
  \bibinfo {author} {\bibfnamefont {S.}~\bibnamefont {Ghosh}},\ }\bibfield
  {title} {\bibinfo {title} {Nonadiabatic modal dynamics around exceptional
  points in an all-lossy dual-mode optical waveguide: Toward chirality-driven
  asymmetric mode conversion},\ }\href
  {https://doi.org/10.1103/PhysRevApplied.10.054008} {\bibfield  {journal}
  {\bibinfo  {journal} {Phys. Rev. Appl.}\ }\textbf {\bibinfo {volume} {10}},\
  \bibinfo {pages} {054008} (\bibinfo {year} {2018})}\BibitemShut {NoStop}%
\bibitem [{\citenamefont {Laha}\ and\ \citenamefont {Ghosh}(2017)}]{Ghosh-17}%
  \BibitemOpen
  \bibfield  {author} {\bibinfo {author} {\bibfnamefont {A.}~\bibnamefont
  {Laha}}\ and\ \bibinfo {author} {\bibfnamefont {S.}~\bibnamefont {Ghosh}},\
  }\bibfield  {title} {\bibinfo {title} {Connected hidden singularities and
  toward successive state flipping in degenerate optical microcavities},\
  }\href {https://doi.org/10.1364/josab.34.000238} {\bibfield  {journal}
  {\bibinfo  {journal} {Journal of the Optical Society of America B}\ }\textbf
  {\bibinfo {volume} {34}},\ \bibinfo {pages} {238} (\bibinfo {year}
  {2017})}\BibitemShut {NoStop}%
\bibitem [{\citenamefont {Laha}\ \emph {et~al.}(2017)\citenamefont {Laha},
  \citenamefont {Biswas},\ and\ \citenamefont {Ghosh}}]{Ghosh-17b}%
  \BibitemOpen
  \bibfield  {author} {\bibinfo {author} {\bibfnamefont {A.}~\bibnamefont
  {Laha}}, \bibinfo {author} {\bibfnamefont {A.}~\bibnamefont {Biswas}},\ and\
  \bibinfo {author} {\bibfnamefont {S.}~\bibnamefont {Ghosh}},\ }\bibfield
  {title} {\bibinfo {title} {Next-nearest-neighbor resonance coupling and
  exceptional singularities in degenerate optical microcavities},\ }\href
  {https://doi.org/10.1364/josab.34.002050} {\bibfield  {journal} {\bibinfo
  {journal} {Journal of the Optical Society of America B}\ }\textbf {\bibinfo
  {volume} {34}},\ \bibinfo {pages} {2050} (\bibinfo {year}
  {2017})}\BibitemShut {NoStop}%
\bibitem [{\citenamefont {Laha}\ \emph {et~al.}(2019)\citenamefont {Laha},
  \citenamefont {Biswas},\ and\ \citenamefont {Ghosh}}]{Ghosh-19}%
  \BibitemOpen
  \bibfield  {author} {\bibinfo {author} {\bibfnamefont {A.}~\bibnamefont
  {Laha}}, \bibinfo {author} {\bibfnamefont {A.}~\bibnamefont {Biswas}},\ and\
  \bibinfo {author} {\bibfnamefont {S.}~\bibnamefont {Ghosh}},\ }\bibfield
  {title} {\bibinfo {title} {Minimally asymmetric state conversion around
  exceptional singularities in a specialty optical microcavity},\ }\href
  {https://doi.org/10.1088/2040-8986/aafd7b} {\bibfield  {journal} {\bibinfo
  {journal} {Journal of Optics}\ }\textbf {\bibinfo {volume} {21}},\ \bibinfo
  {pages} {025201} (\bibinfo {year} {2019})}\BibitemShut {NoStop}%
\bibitem [{\citenamefont {Brandstetter}\ \emph {et~al.}(2014)\citenamefont
  {Brandstetter}, \citenamefont {Liertzer}, \citenamefont {Deutsch},
  \citenamefont {Klang}, \citenamefont {Schöberl}, \citenamefont {Türeci},
  \citenamefont {Strasser}, \citenamefont {Unterrainer},\ and\ \citenamefont
  {Rotter}}]{Rotter-14}%
  \BibitemOpen
  \bibfield  {author} {\bibinfo {author} {\bibfnamefont {M.}~\bibnamefont
  {Brandstetter}}, \bibinfo {author} {\bibfnamefont {M.}~\bibnamefont
  {Liertzer}}, \bibinfo {author} {\bibfnamefont {C.}~\bibnamefont {Deutsch}},
  \bibinfo {author} {\bibfnamefont {P.}~\bibnamefont {Klang}}, \bibinfo
  {author} {\bibfnamefont {J.}~\bibnamefont {Schöberl}}, \bibinfo {author}
  {\bibfnamefont {H.~E.}\ \bibnamefont {Türeci}}, \bibinfo {author}
  {\bibfnamefont {G.}~\bibnamefont {Strasser}}, \bibinfo {author}
  {\bibfnamefont {K.}~\bibnamefont {Unterrainer}},\ and\ \bibinfo {author}
  {\bibfnamefont {S.}~\bibnamefont {Rotter}},\ }\bibfield  {title} {\bibinfo
  {title} {Reversing the pump dependence of a laser at an exceptional point},\
  }\bibfield  {journal} {\bibinfo  {journal} {Nature Communications}\ }\textbf
  {\bibinfo {volume} {5}},\ \href {https://doi.org/10.1038/ncomms5034}
  {10.1038/ncomms5034} (\bibinfo {year} {2014})\BibitemShut {NoStop}%
\bibitem [{\citenamefont {Wong}\ \emph {et~al.}(2016)\citenamefont {Wong},
  \citenamefont {Xu}, \citenamefont {Kim}, \citenamefont
  {O{\textquotesingle}Brien}, \citenamefont {Wang}, \citenamefont {Feng},\ and\
  \citenamefont {Zhang}}]{XZhang-16}%
  \BibitemOpen
  \bibfield  {author} {\bibinfo {author} {\bibfnamefont {Z.~J.}\ \bibnamefont
  {Wong}}, \bibinfo {author} {\bibfnamefont {Y.-L.}\ \bibnamefont {Xu}},
  \bibinfo {author} {\bibfnamefont {J.}~\bibnamefont {Kim}}, \bibinfo {author}
  {\bibfnamefont {K.}~\bibnamefont {O{\textquotesingle}Brien}}, \bibinfo
  {author} {\bibfnamefont {Y.}~\bibnamefont {Wang}}, \bibinfo {author}
  {\bibfnamefont {L.}~\bibnamefont {Feng}},\ and\ \bibinfo {author}
  {\bibfnamefont {X.}~\bibnamefont {Zhang}},\ }\bibfield  {title} {\bibinfo
  {title} {Lasing and anti-lasing in a single cavity},\ }\href
  {https://doi.org/10.1038/nphoton.2016.216} {\bibfield  {journal} {\bibinfo
  {journal} {Nature Photonics}\ }\textbf {\bibinfo {volume} {10}},\ \bibinfo
  {pages} {796} (\bibinfo {year} {2016})}\BibitemShut {NoStop}%
\bibitem [{\citenamefont {Hodaei}\ \emph {et~al.}(2017)\citenamefont {Hodaei},
  \citenamefont {Hassan}, \citenamefont {Wittek}, \citenamefont
  {Garcia-Gracia}, \citenamefont {El-Ganainy}, \citenamefont
  {Christodoulides},\ and\ \citenamefont {Khajavikhan}}]{Khajavikhan-17}%
  \BibitemOpen
  \bibfield  {author} {\bibinfo {author} {\bibfnamefont {H.}~\bibnamefont
  {Hodaei}}, \bibinfo {author} {\bibfnamefont {A.~U.}\ \bibnamefont {Hassan}},
  \bibinfo {author} {\bibfnamefont {S.}~\bibnamefont {Wittek}}, \bibinfo
  {author} {\bibfnamefont {H.}~\bibnamefont {Garcia-Gracia}}, \bibinfo {author}
  {\bibfnamefont {R.}~\bibnamefont {El-Ganainy}}, \bibinfo {author}
  {\bibfnamefont {D.~N.}\ \bibnamefont {Christodoulides}},\ and\ \bibinfo
  {author} {\bibfnamefont {M.}~\bibnamefont {Khajavikhan}},\ }\bibfield
  {title} {\bibinfo {title} {Enhanced sensitivity at higher-order exceptional
  points},\ }\href {https://doi.org/10.1038/nature23280} {\bibfield  {journal}
  {\bibinfo  {journal} {Nature}\ }\textbf {\bibinfo {volume} {548}},\ \bibinfo
  {pages} {187} (\bibinfo {year} {2017})}\BibitemShut {NoStop}%
\bibitem [{\citenamefont {Laha}\ \emph
  {et~al.}(2020{\natexlab{a}})\citenamefont {Laha}, \citenamefont {Beniwal},
  \citenamefont {Dey}, \citenamefont {Biswas},\ and\ \citenamefont
  {Ghosh}}]{Ghosh-20}%
  \BibitemOpen
  \bibfield  {author} {\bibinfo {author} {\bibfnamefont {A.}~\bibnamefont
  {Laha}}, \bibinfo {author} {\bibfnamefont {D.}~\bibnamefont {Beniwal}},
  \bibinfo {author} {\bibfnamefont {S.}~\bibnamefont {Dey}}, \bibinfo {author}
  {\bibfnamefont {A.}~\bibnamefont {Biswas}},\ and\ \bibinfo {author}
  {\bibfnamefont {S.}~\bibnamefont {Ghosh}},\ }\bibfield  {title} {\bibinfo
  {title} {Third-order exceptional point and successive switching among three
  states in an optical microcavity},\ }\href
  {https://doi.org/10.1103/PhysRevA.101.063829} {\bibfield  {journal} {\bibinfo
   {journal} {Phys. Rev. A}\ }\textbf {\bibinfo {volume} {101}},\ \bibinfo
  {pages} {063829} (\bibinfo {year} {2020}{\natexlab{a}})}\BibitemShut
  {NoStop}%
\bibitem [{\citenamefont {Laha}\ \emph {et~al.}(2021)\citenamefont {Laha},
  \citenamefont {Beniwal},\ and\ \citenamefont {Ghosh}}]{Ghosh-21}%
  \BibitemOpen
  \bibfield  {author} {\bibinfo {author} {\bibfnamefont {A.}~\bibnamefont
  {Laha}}, \bibinfo {author} {\bibfnamefont {D.}~\bibnamefont {Beniwal}},\ and\
  \bibinfo {author} {\bibfnamefont {S.}~\bibnamefont {Ghosh}},\ }\bibfield
  {title} {\bibinfo {title} {Successive switching among four states in a
  gain-loss-assisted optical microcavity hosting exceptional points up to order
  four},\ }\href {https://doi.org/10.1103/PhysRevA.103.023526} {\bibfield
  {journal} {\bibinfo  {journal} {Phys. Rev. A}\ }\textbf {\bibinfo {volume}
  {103}},\ \bibinfo {pages} {023526} (\bibinfo {year} {2021})}\BibitemShut
  {NoStop}%
\bibitem [{\citenamefont {Li}\ \emph {et~al.}(2022)\citenamefont {Li},
  \citenamefont {Chen}, \citenamefont {Abbasi}, \citenamefont {Murch},\ and\
  \citenamefont {Whaley}}]{Whaley-22}%
  \BibitemOpen
  \bibfield  {author} {\bibinfo {author} {\bibfnamefont {Z.-Z.}\ \bibnamefont
  {Li}}, \bibinfo {author} {\bibfnamefont {W.}~\bibnamefont {Chen}}, \bibinfo
  {author} {\bibfnamefont {M.}~\bibnamefont {Abbasi}}, \bibinfo {author}
  {\bibfnamefont {K.~W.}\ \bibnamefont {Murch}},\ and\ \bibinfo {author}
  {\bibfnamefont {K.~B.}\ \bibnamefont {Whaley}},\ }\href
  {https://doi.org/10.48550/ARXIV.2210.05048} {\bibinfo {title} {Speeding up
  entanglement generation by proximity to higher-order exceptional points}}
  (\bibinfo {year} {2022})\BibitemShut {NoStop}%
\bibitem [{\citenamefont {Delplace}\ \emph {et~al.}(2021)\citenamefont
  {Delplace}, \citenamefont {Yoshida},\ and\ \citenamefont
  {Hatsugai}}]{Hatsugai-21}%
  \BibitemOpen
  \bibfield  {author} {\bibinfo {author} {\bibfnamefont {P.}~\bibnamefont
  {Delplace}}, \bibinfo {author} {\bibfnamefont {T.}~\bibnamefont {Yoshida}},\
  and\ \bibinfo {author} {\bibfnamefont {Y.}~\bibnamefont {Hatsugai}},\
  }\bibfield  {title} {\bibinfo {title} {Symmetry-protected multifold
  exceptional points and their topological characterization},\ }\href
  {https://doi.org/10.1103/PhysRevLett.127.186602} {\bibfield  {journal}
  {\bibinfo  {journal} {Phys. Rev. Lett.}\ }\textbf {\bibinfo {volume} {127}},\
  \bibinfo {pages} {186602} (\bibinfo {year} {2021})}\BibitemShut {NoStop}%
\bibitem [{\citenamefont {St\aa{}lhammar}\ and\ \citenamefont
  {Bergholtz}(2021)}]{Bergholz-21b}%
  \BibitemOpen
  \bibfield  {author} {\bibinfo {author} {\bibfnamefont {M.}~\bibnamefont
  {St\aa{}lhammar}}\ and\ \bibinfo {author} {\bibfnamefont {E.~J.}\
  \bibnamefont {Bergholtz}},\ }\bibfield  {title} {\bibinfo {title}
  {Classification of exceptional nodal topologies protected by $\mathcal{PT}$
  symmetry},\ }\href {https://doi.org/10.1103/PhysRevB.104.L201104} {\bibfield
  {journal} {\bibinfo  {journal} {Phys. Rev. B}\ }\textbf {\bibinfo {volume}
  {104}},\ \bibinfo {pages} {L201104} (\bibinfo {year} {2021})}\BibitemShut
  {NoStop}%
\bibitem [{\citenamefont {Sayyad}\ and\ \citenamefont
  {Kunst}(2022{\natexlab{a}})}]{Sayyad-22}%
  \BibitemOpen
  \bibfield  {author} {\bibinfo {author} {\bibfnamefont {S.}~\bibnamefont
  {Sayyad}}\ and\ \bibinfo {author} {\bibfnamefont {F.~K.}\ \bibnamefont
  {Kunst}},\ }\bibfield  {title} {\bibinfo {title} {Realizing exceptional
  points of any order in the presence of symmetry},\ }\href
  {https://doi.org/10.1103/PhysRevResearch.4.023130} {\bibfield  {journal}
  {\bibinfo  {journal} {Phys. Rev. Res.}\ }\textbf {\bibinfo {volume} {4}},\
  \bibinfo {pages} {023130} (\bibinfo {year} {2022}{\natexlab{a}})}\BibitemShut
  {NoStop}%
\bibitem [{\citenamefont {Sayyad}\ and\ \citenamefont
  {Kunst}(2022{\natexlab{b}})}]{Kunst-22}%
  \BibitemOpen
  \bibfield  {author} {\bibinfo {author} {\bibfnamefont {S.}~\bibnamefont
  {Sayyad}}\ and\ \bibinfo {author} {\bibfnamefont {F.~K.}\ \bibnamefont
  {Kunst}},\ }\bibfield  {title} {\bibinfo {title} {Realizing exceptional
  points of any order in the presence of symmetry},\ }\href
  {https://doi.org/10.1103/PhysRevResearch.4.023130} {\bibfield  {journal}
  {\bibinfo  {journal} {Phys. Rev. Res.}\ }\textbf {\bibinfo {volume} {4}},\
  \bibinfo {pages} {023130} (\bibinfo {year} {2022}{\natexlab{b}})}\BibitemShut
  {NoStop}%
\bibitem [{\citenamefont
  {Mostafazadeh}(2002{\natexlab{a}})}]{mostafazadeh2002pseudo}%
  \BibitemOpen
  \bibfield  {author} {\bibinfo {author} {\bibfnamefont {A.}~\bibnamefont
  {Mostafazadeh}},\ }\bibfield  {title} {\bibinfo {title} {Pseudo-hermiticity
  versus pt symmetry: the necessary condition for the reality of the spectrum
  of a non-hermitian hamiltonian},\ }\href@noop {} {\bibfield  {journal}
  {\bibinfo  {journal} {Journal of Mathematical Physics}\ }\textbf {\bibinfo
  {volume} {43}},\ \bibinfo {pages} {205} (\bibinfo {year}
  {2002}{\natexlab{a}})}\BibitemShut {NoStop}%
\bibitem [{\citenamefont {Mostafazadeh}(2002{\natexlab{b}})}]{mostafazadeh2}%
  \BibitemOpen
  \bibfield  {author} {\bibinfo {author} {\bibfnamefont {A.}~\bibnamefont
  {Mostafazadeh}},\ }\bibfield  {title} {\bibinfo {title} {Pseudo-hermiticity
  versus pt-symmetry. ii. a complete characterization of non-hermitian
  hamiltonians with a real spectrum},\ }\href@noop {} {\bibfield  {journal}
  {\bibinfo  {journal} {Journal of Mathematical Physics}\ }\textbf {\bibinfo
  {volume} {43}},\ \bibinfo {pages} {2814} (\bibinfo {year}
  {2002}{\natexlab{b}})}\BibitemShut {NoStop}%
\bibitem [{\citenamefont {Mostafazadeh}(2002{\natexlab{c}})}]{mostafazadeh3}%
  \BibitemOpen
  \bibfield  {author} {\bibinfo {author} {\bibfnamefont {A.}~\bibnamefont
  {Mostafazadeh}},\ }\bibfield  {title} {\bibinfo {title} {Pseudo-hermiticity
  versus pt-symmetry iii: Equivalence of pseudo-hermiticity and the presence of
  antilinear symmetries},\ }\href@noop {} {\bibfield  {journal} {\bibinfo
  {journal} {Journal of Mathematical Physics}\ }\textbf {\bibinfo {volume}
  {43}},\ \bibinfo {pages} {3944} (\bibinfo {year}
  {2002}{\natexlab{c}})}\BibitemShut {NoStop}%
\bibitem [{\citenamefont
  {Mostafazadeh}(2010{\natexlab{a}})}]{mostafazadeh2010}%
  \BibitemOpen
  \bibfield  {author} {\bibinfo {author} {\bibfnamefont {A.}~\bibnamefont
  {Mostafazadeh}},\ }\bibfield  {title} {\bibinfo {title} {Pseudo-hermitian
  representation of quantum mechanics},\ }\href@noop {} {\bibfield  {journal}
  {\bibinfo  {journal} {International Journal of Geometric Methods in Modern
  Physics}\ }\textbf {\bibinfo {volume} {7}},\ \bibinfo {pages} {1191}
  (\bibinfo {year} {2010}{\natexlab{a}})}\BibitemShut {NoStop}%
\bibitem [{\citenamefont {Ashida}\ \emph {et~al.}(2020)\citenamefont {Ashida},
  \citenamefont {Gong},\ and\ \citenamefont {Ueda}}]{Ashida-20}%
  \BibitemOpen
  \bibfield  {author} {\bibinfo {author} {\bibfnamefont {Y.}~\bibnamefont
  {Ashida}}, \bibinfo {author} {\bibfnamefont {Z.}~\bibnamefont {Gong}},\ and\
  \bibinfo {author} {\bibfnamefont {M.}~\bibnamefont {Ueda}},\ }\bibfield
  {title} {\bibinfo {title} {Non-hermitian physics},\ }\href
  {https://doi.org/10.1080/00018732.2021.1876991} {\bibfield  {journal}
  {\bibinfo  {journal} {Advances in Physics}\ }\textbf {\bibinfo {volume}
  {69}},\ \bibinfo {pages} {249} (\bibinfo {year} {2020})}\BibitemShut
  {NoStop}%
\bibitem [{\citenamefont {Zhang}\ \emph {et~al.}(2020)\citenamefont {Zhang},
  \citenamefont {Qin},\ and\ \citenamefont {Xiao}}]{ZhangQinXiao-20}%
  \BibitemOpen
  \bibfield  {author} {\bibinfo {author} {\bibfnamefont {R.}~\bibnamefont
  {Zhang}}, \bibinfo {author} {\bibfnamefont {H.}~\bibnamefont {Qin}},\ and\
  \bibinfo {author} {\bibfnamefont {J.}~\bibnamefont {Xiao}},\ }\bibfield
  {title} {\bibinfo {title} {{PT}-symmetry entails pseudo-hermiticity
  regardless of diagonalizability},\ }\href {https://doi.org/10.1063/1.5117211}
  {\bibfield  {journal} {\bibinfo  {journal} {Journal of Mathematical Physics}\
  }\textbf {\bibinfo {volume} {61}},\ \bibinfo {pages} {012101} (\bibinfo
  {year} {2020})}\BibitemShut {NoStop}%
\bibitem [{\citenamefont {Bian}\ \emph {et~al.}(2020)\citenamefont {Bian},
  \citenamefont {Xiao}, \citenamefont {Wang}, \citenamefont {Zhan},
  \citenamefont {Onanga}, \citenamefont {Ruzicka}, \citenamefont {Yi},
  \citenamefont {Joglekar},\ and\ \citenamefont {Xue}}]{bian2020conserved}%
  \BibitemOpen
  \bibfield  {author} {\bibinfo {author} {\bibfnamefont {Z.}~\bibnamefont
  {Bian}}, \bibinfo {author} {\bibfnamefont {L.}~\bibnamefont {Xiao}}, \bibinfo
  {author} {\bibfnamefont {K.}~\bibnamefont {Wang}}, \bibinfo {author}
  {\bibfnamefont {X.}~\bibnamefont {Zhan}}, \bibinfo {author} {\bibfnamefont
  {F.~A.}\ \bibnamefont {Onanga}}, \bibinfo {author} {\bibfnamefont
  {F.}~\bibnamefont {Ruzicka}}, \bibinfo {author} {\bibfnamefont
  {W.}~\bibnamefont {Yi}}, \bibinfo {author} {\bibfnamefont {Y.~N.}\
  \bibnamefont {Joglekar}},\ and\ \bibinfo {author} {\bibfnamefont
  {P.}~\bibnamefont {Xue}},\ }\bibfield  {title} {\bibinfo {title} {Conserved
  quantities in parity-time symmetric systems},\ }\href@noop {} {\bibfield
  {journal} {\bibinfo  {journal} {Physical Review Research}\ }\textbf {\bibinfo
  {volume} {2}},\ \bibinfo {pages} {022039} (\bibinfo {year}
  {2020})}\BibitemShut {NoStop}%
\bibitem [{\citenamefont {Agarwal}\ \emph {et~al.}(2022)\citenamefont
  {Agarwal}, \citenamefont {Muldoon},\ and\ \citenamefont
  {Joglekar}}]{agarwal2022conserved}%
  \BibitemOpen
  \bibfield  {author} {\bibinfo {author} {\bibfnamefont {K.~S.}\ \bibnamefont
  {Agarwal}}, \bibinfo {author} {\bibfnamefont {J.}~\bibnamefont {Muldoon}},\
  and\ \bibinfo {author} {\bibfnamefont {Y.~N.}\ \bibnamefont {Joglekar}},\
  }\bibfield  {title} {\bibinfo {title} {Conserved quantities in non-hermitian
  systems via vectorization method},\ }\href@noop {} {\bibfield  {journal}
  {\bibinfo  {journal} {arXiv preprint arXiv:2201.05019}\ } (\bibinfo {year}
  {2022})}\BibitemShut {NoStop}%
\bibitem [{\citenamefont {Starkov}\ \emph {et~al.}(2022)\citenamefont
  {Starkov}, \citenamefont {Fistoul},\ and\ \citenamefont
  {Eremin}}]{StarkovFistulEremin-22}%
  \BibitemOpen
  \bibfield  {author} {\bibinfo {author} {\bibfnamefont {G.~A.}\ \bibnamefont
  {Starkov}}, \bibinfo {author} {\bibfnamefont {M.~V.}\ \bibnamefont
  {Fistoul}},\ and\ \bibinfo {author} {\bibfnamefont {I.~M.}\ \bibnamefont
  {Eremin}},\ }\href {https://doi.org/10.48550/ARXIV.2211.00679} {\bibinfo
  {title} {Quantum phase transitions in non-hermitian
  $\mathcal{P}\mathcal{T}$-symmetric transverse-field ising spin chains}}
  (\bibinfo {year} {2022})\BibitemShut {NoStop}%
\bibitem [{\citenamefont {Lieb}\ \emph {et~al.}(1961)\citenamefont {Lieb},
  \citenamefont {Schultz},\ and\ \citenamefont {Mattis}}]{Lieb-61}%
  \BibitemOpen
  \bibfield  {author} {\bibinfo {author} {\bibfnamefont {E.}~\bibnamefont
  {Lieb}}, \bibinfo {author} {\bibfnamefont {T.}~\bibnamefont {Schultz}},\ and\
  \bibinfo {author} {\bibfnamefont {D.}~\bibnamefont {Mattis}},\ }\bibfield
  {title} {\bibinfo {title} {Two soluble models of an antiferromagnetic
  chain},\ }\href {https://doi.org/10.1016/0003-4916(61)90115-4} {\bibfield
  {journal} {\bibinfo  {journal} {Annals of Physics}\ }\textbf {\bibinfo
  {volume} {16}},\ \bibinfo {pages} {407} (\bibinfo {year} {1961})}\BibitemShut
  {NoStop}%
\bibitem [{\citenamefont {Ding}\ \emph {et~al.}(2021)\citenamefont {Ding},
  \citenamefont {Shi}, \citenamefont {Zhang}, \citenamefont {Shen},
  \citenamefont {Zhang},\ and\ \citenamefont {Zhang}}]{ding2021experimental}%
  \BibitemOpen
  \bibfield  {author} {\bibinfo {author} {\bibfnamefont {L.}~\bibnamefont
  {Ding}}, \bibinfo {author} {\bibfnamefont {K.}~\bibnamefont {Shi}}, \bibinfo
  {author} {\bibfnamefont {Q.}~\bibnamefont {Zhang}}, \bibinfo {author}
  {\bibfnamefont {D.}~\bibnamefont {Shen}}, \bibinfo {author} {\bibfnamefont
  {X.}~\bibnamefont {Zhang}},\ and\ \bibinfo {author} {\bibfnamefont
  {W.}~\bibnamefont {Zhang}},\ }\bibfield  {title} {\bibinfo {title}
  {Experimental determination of p t-symmetric exceptional points in a single
  trapped ion},\ }\href@noop {} {\bibfield  {journal} {\bibinfo  {journal}
  {Physical Review Letters}\ }\textbf {\bibinfo {volume} {126}},\ \bibinfo
  {pages} {083604} (\bibinfo {year} {2021})}\BibitemShut {NoStop}%
\bibitem [{\citenamefont {Louren{\c{c}}o}\ \emph {et~al.}(2022)\citenamefont
  {Louren{\c{c}}o}, \citenamefont {Higgins}, \citenamefont {Zhang},
  \citenamefont {Hennrich},\ and\ \citenamefont
  {Macr{\`\i}}}]{lourencco2022non}%
  \BibitemOpen
  \bibfield  {author} {\bibinfo {author} {\bibfnamefont {J.~A.}\ \bibnamefont
  {Louren{\c{c}}o}}, \bibinfo {author} {\bibfnamefont {G.}~\bibnamefont
  {Higgins}}, \bibinfo {author} {\bibfnamefont {C.}~\bibnamefont {Zhang}},
  \bibinfo {author} {\bibfnamefont {M.}~\bibnamefont {Hennrich}},\ and\
  \bibinfo {author} {\bibfnamefont {T.}~\bibnamefont {Macr{\`\i}}},\ }\bibfield
   {title} {\bibinfo {title} {Non-hermitian dynamics and pt-symmetry breaking
  in interacting mesoscopic rydberg platforms},\ }\href@noop {} {\bibfield
  {journal} {\bibinfo  {journal} {Physical Review A}\ }\textbf {\bibinfo
  {volume} {106}},\ \bibinfo {pages} {023309} (\bibinfo {year}
  {2022})}\BibitemShut {NoStop}%
\bibitem [{\citenamefont {Li}\ \emph {et~al.}(2019)\citenamefont {Li},
  \citenamefont {Harter}, \citenamefont {Liu}, \citenamefont {de~Melo},
  \citenamefont {Joglekar},\ and\ \citenamefont {Luo}}]{li2019observation}%
  \BibitemOpen
  \bibfield  {author} {\bibinfo {author} {\bibfnamefont {J.}~\bibnamefont
  {Li}}, \bibinfo {author} {\bibfnamefont {A.~K.}\ \bibnamefont {Harter}},
  \bibinfo {author} {\bibfnamefont {J.}~\bibnamefont {Liu}}, \bibinfo {author}
  {\bibfnamefont {L.}~\bibnamefont {de~Melo}}, \bibinfo {author} {\bibfnamefont
  {Y.~N.}\ \bibnamefont {Joglekar}},\ and\ \bibinfo {author} {\bibfnamefont
  {L.}~\bibnamefont {Luo}},\ }\bibfield  {title} {\bibinfo {title} {Observation
  of parity-time symmetry breaking transitions in a dissipative floquet system
  of ultracold atoms},\ }\href@noop {} {\bibfield  {journal} {\bibinfo
  {journal} {Nature communications}\ }\textbf {\bibinfo {volume} {10}},\
  \bibinfo {pages} {1} (\bibinfo {year} {2019})}\BibitemShut {NoStop}%
\bibitem [{\citenamefont {Cartarius}\ and\ \citenamefont
  {Wunner}(2012)}]{cartarius2012model}%
  \BibitemOpen
  \bibfield  {author} {\bibinfo {author} {\bibfnamefont {H.}~\bibnamefont
  {Cartarius}}\ and\ \bibinfo {author} {\bibfnamefont {G.}~\bibnamefont
  {Wunner}},\ }\bibfield  {title} {\bibinfo {title} {Model of a pt-symmetric
  bose-einstein condensate in a $\delta$-function double-well potential},\
  }\href@noop {} {\bibfield  {journal} {\bibinfo  {journal} {Physical Review
  A}\ }\textbf {\bibinfo {volume} {86}},\ \bibinfo {pages} {013612} (\bibinfo
  {year} {2012})}\BibitemShut {NoStop}%
\bibitem [{\citenamefont {Naghiloo}\ \emph {et~al.}(2019)\citenamefont
  {Naghiloo}, \citenamefont {Abbasi}, \citenamefont {Joglekar},\ and\
  \citenamefont {Murch}}]{naghiloo2019}%
  \BibitemOpen
  \bibfield  {author} {\bibinfo {author} {\bibfnamefont {M.}~\bibnamefont
  {Naghiloo}}, \bibinfo {author} {\bibfnamefont {M.}~\bibnamefont {Abbasi}},
  \bibinfo {author} {\bibfnamefont {Y.~N.}\ \bibnamefont {Joglekar}},\ and\
  \bibinfo {author} {\bibfnamefont {K.}~\bibnamefont {Murch}},\ }\bibfield
  {title} {\bibinfo {title} {Quantum state tomography across the exceptional
  point in a single dissipative qubit},\ }\href@noop {} {\bibfield  {journal}
  {\bibinfo  {journal} {Nature Physics}\ }\textbf {\bibinfo {volume} {15}},\
  \bibinfo {pages} {1232} (\bibinfo {year} {2019})}\BibitemShut {NoStop}%
\bibitem [{\citenamefont {Dogra}\ \emph {et~al.}(2021)\citenamefont {Dogra},
  \citenamefont {Melnikov},\ and\ \citenamefont {Paraoanu}}]{dogra2021quantum}%
  \BibitemOpen
  \bibfield  {author} {\bibinfo {author} {\bibfnamefont {S.}~\bibnamefont
  {Dogra}}, \bibinfo {author} {\bibfnamefont {A.~A.}\ \bibnamefont
  {Melnikov}},\ and\ \bibinfo {author} {\bibfnamefont {G.~S.}\ \bibnamefont
  {Paraoanu}},\ }\bibfield  {title} {\bibinfo {title} {Quantum simulation of
  parity--time symmetry breaking with a superconducting quantum processor},\
  }\href@noop {} {\bibfield  {journal} {\bibinfo  {journal} {Communications
  Physics}\ }\textbf {\bibinfo {volume} {4}},\ \bibinfo {pages} {1} (\bibinfo
  {year} {2021})}\BibitemShut {NoStop}%
\bibitem [{\citenamefont {Wu}\ \emph {et~al.}(2019)\citenamefont {Wu},
  \citenamefont {Liu}, \citenamefont {Geng}, \citenamefont {Song},
  \citenamefont {Ye}, \citenamefont {Duan}, \citenamefont {Rong},\ and\
  \citenamefont {Du}}]{wu2019observation}%
  \BibitemOpen
  \bibfield  {author} {\bibinfo {author} {\bibfnamefont {Y.}~\bibnamefont
  {Wu}}, \bibinfo {author} {\bibfnamefont {W.}~\bibnamefont {Liu}}, \bibinfo
  {author} {\bibfnamefont {J.}~\bibnamefont {Geng}}, \bibinfo {author}
  {\bibfnamefont {X.}~\bibnamefont {Song}}, \bibinfo {author} {\bibfnamefont
  {X.}~\bibnamefont {Ye}}, \bibinfo {author} {\bibfnamefont {C.-K.}\
  \bibnamefont {Duan}}, \bibinfo {author} {\bibfnamefont {X.}~\bibnamefont
  {Rong}},\ and\ \bibinfo {author} {\bibfnamefont {J.}~\bibnamefont {Du}},\
  }\bibfield  {title} {\bibinfo {title} {Observation of parity-time symmetry
  breaking in a single-spin system},\ }\href@noop {} {\bibfield  {journal}
  {\bibinfo  {journal} {Science}\ }\textbf {\bibinfo {volume} {364}},\ \bibinfo
  {pages} {878} (\bibinfo {year} {2019})}\BibitemShut {NoStop}%
\bibitem [{\citenamefont {Tetling}\ \emph {et~al.}(2022)\citenamefont
  {Tetling}, \citenamefont {Fistul},\ and\ \citenamefont
  {Eremin}}]{tetling2022linear}%
  \BibitemOpen
  \bibfield  {author} {\bibinfo {author} {\bibfnamefont {L.}~\bibnamefont
  {Tetling}}, \bibinfo {author} {\bibfnamefont {M.}~\bibnamefont {Fistul}},\
  and\ \bibinfo {author} {\bibfnamefont {I.~M.}\ \bibnamefont {Eremin}},\
  }\bibfield  {title} {\bibinfo {title} {Linear response for pseudo-hermitian
  hamiltonian systems: Application to pt-symmetric qubits},\ }\href@noop {}
  {\bibfield  {journal} {\bibinfo  {journal} {Physical Review B}\ }\textbf
  {\bibinfo {volume} {106}},\ \bibinfo {pages} {134511} (\bibinfo {year}
  {2022})}\BibitemShut {NoStop}%
\bibitem [{\citenamefont {Mostafazadeh}(2010{\natexlab{b}})}]{Mostafazadeh-10}%
  \BibitemOpen
  \bibfield  {author} {\bibinfo {author} {\bibfnamefont {A.}~\bibnamefont
  {Mostafazadeh}},\ }\bibfield  {title} {\bibinfo {title} {{Pseudo}-{Hermitian}
  {Representation} {of} {Quantum} {Mechanics}},\ }\href
  {https://doi.org/10.1142/s0219887810004816} {\bibfield  {journal} {\bibinfo
  {journal} {International Journal of Geometric Methods in Modern Physics}\
  }\textbf {\bibinfo {volume} {07}},\ \bibinfo {pages} {1191} (\bibinfo {year}
  {2010}{\natexlab{b}})}\BibitemShut {NoStop}%
\bibitem [{\citenamefont {Dirac}(1930)}]{Dirac-1930}%
  \BibitemOpen
  \bibfield  {author} {\bibinfo {author} {\bibfnamefont {P.~A.~M.}\
  \bibnamefont {Dirac}},\ }\href@noop {} {\emph {\bibinfo {title} {The
  Principles of Quantum Mechanics}}}\ (\bibinfo  {publisher} {Clarendon
  Press},\ \bibinfo {year} {1930})\BibitemShut {NoStop}%
\bibitem [{\citenamefont {Li}\ \emph {et~al.}(2014)\citenamefont {Li},
  \citenamefont {Zhang}, \citenamefont {Zhang},\ and\ \citenamefont
  {Song}}]{Song-14}%
  \BibitemOpen
  \bibfield  {author} {\bibinfo {author} {\bibfnamefont {C.}~\bibnamefont
  {Li}}, \bibinfo {author} {\bibfnamefont {G.}~\bibnamefont {Zhang}}, \bibinfo
  {author} {\bibfnamefont {X.~Z.}\ \bibnamefont {Zhang}},\ and\ \bibinfo
  {author} {\bibfnamefont {Z.}~\bibnamefont {Song}},\ }\bibfield  {title}
  {\bibinfo {title} {Conventional quantum phase transition driven by a complex
  parameter in a non-hermitian $\mathcal{PT}\ensuremath{-}\mathrm{symmetric}$
  ising model},\ }\href {https://doi.org/10.1103/PhysRevA.90.012103} {\bibfield
   {journal} {\bibinfo  {journal} {Phys. Rev. A}\ }\textbf {\bibinfo {volume}
  {90}},\ \bibinfo {pages} {012103} (\bibinfo {year} {2014})}\BibitemShut
  {NoStop}%
\bibitem [{\citenamefont {Li}\ and\ \citenamefont {Song}(2015)}]{Song-15}%
  \BibitemOpen
  \bibfield  {author} {\bibinfo {author} {\bibfnamefont {C.}~\bibnamefont
  {Li}}\ and\ \bibinfo {author} {\bibfnamefont {Z.}~\bibnamefont {Song}},\
  }\bibfield  {title} {\bibinfo {title} {Finite-temperature quantum criticality
  in a complex-parameter plane},\ }\href
  {https://doi.org/10.1103/PhysRevA.92.062103} {\bibfield  {journal} {\bibinfo
  {journal} {Phys. Rev. A}\ }\textbf {\bibinfo {volume} {92}},\ \bibinfo
  {pages} {062103} (\bibinfo {year} {2015})}\BibitemShut {NoStop}%
\bibitem [{\citenamefont {Lenke}\ \emph {et~al.}(2021)\citenamefont {Lenke},
  \citenamefont {M\"uhlhauser},\ and\ \citenamefont {Schmidt}}]{Schmidt-21}%
  \BibitemOpen
  \bibfield  {author} {\bibinfo {author} {\bibfnamefont {L.}~\bibnamefont
  {Lenke}}, \bibinfo {author} {\bibfnamefont {M.}~\bibnamefont
  {M\"uhlhauser}},\ and\ \bibinfo {author} {\bibfnamefont {K.~P.}\ \bibnamefont
  {Schmidt}},\ }\bibfield  {title} {\bibinfo {title} {High-order series
  expansion of non-hermitian quantum spin models},\ }\href
  {https://doi.org/10.1103/PhysRevB.104.195137} {\bibfield  {journal} {\bibinfo
   {journal} {Phys. Rev. B}\ }\textbf {\bibinfo {volume} {104}},\ \bibinfo
  {pages} {195137} (\bibinfo {year} {2021})}\BibitemShut {NoStop}%
\bibitem [{\citenamefont {Neumann}\ and\ \citenamefont
  {Wigner}(1929)}]{NeumannWigner-29}%
  \BibitemOpen
  \bibfield  {author} {\bibinfo {author} {\bibfnamefont {J.~v.}\ \bibnamefont
  {Neumann}}\ and\ \bibinfo {author} {\bibfnamefont {E.}~\bibnamefont
  {Wigner}},\ }\bibfield  {title} {\bibinfo {title} {Über das verhalten von
  eigenwerten bei adiabatischen prozessen},\ }\href@noop {} {\bibfield
  {journal} {\bibinfo  {journal} {Physikalische Zeitschrift}\ }\textbf
  {\bibinfo {volume} {30}},\ \bibinfo {pages} {467} (\bibinfo {year}
  {1929})}\BibitemShut {NoStop}%
\bibitem [{\citenamefont {Grifoni}\ and\ \citenamefont
  {H{\"a}nggi}(1998)}]{grifoni1998driven}%
  \BibitemOpen
  \bibfield  {author} {\bibinfo {author} {\bibfnamefont {M.}~\bibnamefont
  {Grifoni}}\ and\ \bibinfo {author} {\bibfnamefont {P.}~\bibnamefont
  {H{\"a}nggi}},\ }\bibfield  {title} {\bibinfo {title} {Driven quantum
  tunneling},\ }\href@noop {} {\bibfield  {journal} {\bibinfo  {journal}
  {Physics Reports}\ }\textbf {\bibinfo {volume} {304}},\ \bibinfo {pages}
  {229} (\bibinfo {year} {1998})}\BibitemShut {NoStop}%
\bibitem [{\citenamefont {Laha}\ \emph
  {et~al.}(2020{\natexlab{b}})\citenamefont {Laha}, \citenamefont {Beniwal},
  \citenamefont {Dey}, \citenamefont {Biswas},\ and\ \citenamefont
  {Ghosh}}]{Somnath-20}%
  \BibitemOpen
  \bibfield  {author} {\bibinfo {author} {\bibfnamefont {A.}~\bibnamefont
  {Laha}}, \bibinfo {author} {\bibfnamefont {D.}~\bibnamefont {Beniwal}},
  \bibinfo {author} {\bibfnamefont {S.}~\bibnamefont {Dey}}, \bibinfo {author}
  {\bibfnamefont {A.}~\bibnamefont {Biswas}},\ and\ \bibinfo {author}
  {\bibfnamefont {S.}~\bibnamefont {Ghosh}},\ }\bibfield  {title} {\bibinfo
  {title} {Third-order exceptional point and successive switching among three
  states in an optical microcavity},\ }\href
  {https://doi.org/10.1103/PhysRevA.101.063829} {\bibfield  {journal} {\bibinfo
   {journal} {Phys. Rev. A}\ }\textbf {\bibinfo {volume} {101}},\ \bibinfo
  {pages} {063829} (\bibinfo {year} {2020}{\natexlab{b}})}\BibitemShut
  {NoStop}%
\bibitem [{\citenamefont {Mandal}\ and\ \citenamefont
  {Bergholtz}(2021)}]{Bergholtz-21}%
  \BibitemOpen
  \bibfield  {author} {\bibinfo {author} {\bibfnamefont {I.}~\bibnamefont
  {Mandal}}\ and\ \bibinfo {author} {\bibfnamefont {E.~J.}\ \bibnamefont
  {Bergholtz}},\ }\bibfield  {title} {\bibinfo {title} {Symmetry and
  higher-order exceptional points},\ }\href
  {https://doi.org/10.1103/PhysRevLett.127.186601} {\bibfield  {journal}
  {\bibinfo  {journal} {Phys. Rev. Lett.}\ }\textbf {\bibinfo {volume} {127}},\
  \bibinfo {pages} {186601} (\bibinfo {year} {2021})}\BibitemShut {NoStop}%
\bibitem [{\citenamefont {Krein}(1950)}]{Krein-1950}%
  \BibitemOpen
  \bibfield  {author} {\bibinfo {author} {\bibfnamefont {M.}~\bibnamefont
  {Krein}},\ }\bibfield  {title} {\bibinfo {title} {A generalization of some
  investigations of linear differential equations with periodic coefficients},\
  }in\ \href@noop {} {\emph {\bibinfo {booktitle} {Doklady Akad. Nauk SSSR}}},\
  Vol.~\bibinfo {volume} {73}\ (\bibinfo {year} {1950})\ pp.\ \bibinfo {pages}
  {445--448}\BibitemShut {NoStop}%
\bibitem [{\citenamefont {Gel'fand}\ and\ \citenamefont
  {Lidskii}(1955)}]{GelfandLidskii-1955}%
  \BibitemOpen
  \bibfield  {author} {\bibinfo {author} {\bibfnamefont {I.~M.}\ \bibnamefont
  {Gel'fand}}\ and\ \bibinfo {author} {\bibfnamefont {V.~B.}\ \bibnamefont
  {Lidskii}},\ }\bibfield  {title} {\bibinfo {title} {On the structure of the
  regions of stability of linear canonical systems of differential equations
  with periodic coefficients},\ }\href@noop {} {\bibfield  {journal} {\bibinfo
  {journal} {Uspekhi Matematicheskikh Nauk}\ }\textbf {\bibinfo {volume}
  {10}},\ \bibinfo {pages} {3} (\bibinfo {year} {1955})}\BibitemShut {NoStop}%
\bibitem [{\citenamefont {Starzhinskii}\ and\ \citenamefont
  {Yakubovich}(1975)}]{StarzhinskiiYakubovich-1975}%
  \BibitemOpen
  \bibfield  {author} {\bibinfo {author} {\bibfnamefont {V.}~\bibnamefont
  {Starzhinskii}}\ and\ \bibinfo {author} {\bibfnamefont {V.}~\bibnamefont
  {Yakubovich}},\ }\href@noop {} {\emph {\bibinfo {title} {Linear differential
  equations with periodic coefficients 2 vol.}}}\ (\bibinfo  {publisher} {Wiley
  London},\ \bibinfo {year} {1975})\BibitemShut {NoStop}%
\end{thebibliography}

%
 
\end{document}